\documentclass[fleqn,usenatbib]{stile/mnras}

\usepackage{txfonts}
\usepackage[T1]{fontenc}
\usepackage{ae,aecompl}

\usepackage{graphicx}

\usepackage{amsmath}
\usepackage{amsfonts}
\usepackage{amssymb}

\usepackage{xcolor}

\usepackage{breakcites}
\usepackage{hyperref}
\usepackage{footnote}

\usepackage{comment}

\usepackage{xfrac}        

\def\nat{Nature}

\def\mnras{MNRAS}
\def\apj{ApJ}
\def\apjl{ApJL}
\def\apjs{ApJS}
\def\aap{A\&A}

\def\araa{ARA\&A}
\def\aj{AJ}

\def\physrep{Phys. Rep.}

\def\pasp{Publ. Astr. Soc. Pac.}

\def\pasj{Pub. Astron. Soc. Japan}

\newcommand{\quotes}[1]{``#1''}

\def\gtsima{$\; \buildrel > \over \sim \;$}
\def\ltsima{$\; \buildrel < \over \sim \;$}
\def\gsim{\lower.5ex\hbox{\gtsima}}
\def\lsim{\lower.5ex\hbox{\ltsima}}
\def\gtsima{$\; \buildrel > \over \sim \;$} 
\def\ltsima{$\; \buildrel < \over \sim \;$} \def\gsim{\lower.5ex\hbox{\gtsima}} 
\def\lsim{\lower.5ex\hbox{\ltsima}} 
\def\simgt{\lower.5ex\hbox{\gtsima}} 
\def\simlt{\lower.5ex\hbox{\ltsima}}

\def\be{\begin{equation}}
\def\ee{\end{equation}}

\def\NII{\hbox{[N~$\scriptstyle\rm II $]}}
\def\HII{\hbox{H~$\scriptstyle\rm II $~}}
\def\HI{\hbox{H~$\scriptstyle\rm I $~}}
\def\Hb{\hbox{H~$ \beta $~}}
\def\Ha{\hbox{H~$ \alpha $~}}
\def\OI{\hbox{[O~$\scriptstyle\rm I $]}}
\def\OII{\hbox{[O~$\scriptstyle\rm II $]}}
\def\CII{\hbox{[C~$\scriptstyle\rm II $]}}
\def\HI{\hbox{H~$\scriptstyle\rm I\ $}} 
\def\HII{\hbox{H~$\scriptstyle\rm II\ $}}

\def\NIII{\hbox{[N~$\scriptstyle\rm III $]}}
\def\OIII{\hbox{[O~$\scriptstyle\rm III $]}}
\def\CIII{\hbox{C~$\scriptstyle\rm III $]}}

\graphicspath{{plots/}}                 

\begin{document}

\definecolor{apcolor}{HTML}{b3003b}
\definecolor{color}{HTML}{b1331a}

\defcitealias{F16}{F17}
\defcitealias{weingartner2001dust}{WD01}
\def\WD01{\citetalias{weingartner2001dust}}
\def\F17{\citetalias{F16}}
\def\C15{\citetalias{capak15}}

\title[Dust temperature in high-$z$ galaxies]{Dust temperature in ALMA \CII -detected high-$z$ galaxies}

\author[Sommovigo et al.]{L. Sommovigo$^{1}$\thanks{\href{mailto:laura.sommovigo@sns.it}{laura.sommovigo@sns.it}},
A. Ferrara$^{1}$, S. Carniani$^{1}$, A. Zanella$^{2}$, A. Pallottini$^{1}$, S. Gallerani$^{1}$,
\newauthor L. Vallini$^{1}$\\
$^{1}$ Scuola Normale Superiore, Piazza dei Cavalieri 7, I-56126 Pisa, Italy\\
$^2$ INAF - Osservatorio Astronomico di Padova, Vicolo Osservatorio 5,  I-35122 Padova
}

\label{firstpage}
\pagerange{\pageref{firstpage}--\pageref{lastpage}}

\maketitle

\begin{abstract}
At redshift $z>5$ the far-infrared (FIR) continuum spectra of main-sequence galaxies are sparsely sampled, often with a single data point. The dust temperature $T_{\rm d, SED}$ thus has to be assumed in the FIR continuum fitting. This introduces large uncertainties regarding the derived dust mass ($M_{\rm d}$), FIR luminosity, and obscured fraction of the star formation rate. These are crucial quantities to quantify the effect of dust obscuration in high-$z$ galaxies. To overcome observations limitations, we introduce a new method that combines dust continuum information with the overlying \CII$158\mu$m line emission. By breaking the $M_{\rm d} - T_{\rm d, SED}$ degeneracy, with our method, we can reliably constrain the dust temperature with a single observation at $158\mu$m. This method can be applied to all ALMA and NOEMA \CII\ observations, and exploited in ALMA Large Programs such as ALPINE and REBELS targeting \CII\ emitters at high-$z$. We also provide a physical interpretation of the empirical relation recently found between \textit{molecular} gas mass and \CII\ luminosity. We derive an analogous relation linking the \textit{total} gas surface density and \CII\ surface brightness. By combining the two, we predict the cosmic evolution of the surface density ratio $\Sigma_{\rm H_2} / \Sigma_{\rm gas}$. We find that $\Sigma_{\rm H_2} / \Sigma_{\rm gas}$ slowly increases with redshift, which is compatible with current observations at $0 < z < 4$. 
\end{abstract}


\section{Introduction}\label{intro}

The Hubble Space Telescope (HST) and ground-based telescopes have been used to investigate the rest-frame Ultraviolet (UV) emission from early galaxies (for a recent theoretical review see \citealt{Dayal18}). The advent of high sensitivity millimetre interferometers such as the Atacama Large Millimeter Array (ALMA), allowed us for the first time to study also the Far-Infrared (FIR) emission from these sources \citep[see e.g.][]{2013carilli}. 

ALMA can detect both the FIR continuum and the brightest FIR lines in \quotes{normal} (i.e. main sequence) galaxies at $z\ge 4$ \citep[see e.g.][]{capak15,willott2015star,Bouwens16,laporte:2017apj,barisic2017dust,carniani:2017oiii,bowler2018obscured,carniani2018alma,carniani2018clumps,2019A&A...631A.167D,Tamura19,Bakx20,bethermin_alpine,Schaerer_alpine}. The FIR continuum is emitted as thermal radiation by dust grains, heated through the absorption of UV and optical light from newly born stars \citep[see e.g.][]{draine1989interstellar,meurer1999dust,calzetti2000dust,weingartner2001dust,Draine03}. 

The galaxy properties that mainly characterise the FIR continuum emission are the dust temperature $T_{\rm d, SED}$ and the dust mass $M_{\rm d}$, which are degenerate quantities. For the simultaneous determination of $T_{\rm d, SED}$ and $M_{\rm d}$ the most common approach is to fit the observed Spectral Energy Distribution (SED) with a single temperature\footnote{We underline that $T_{\rm d, SED}$ does not necessarily correspond to the dust physical temperature, which is instead characterised by a Probability Distribution Function \citep[PDF, see e.g.][]{behrens2018dusty,nostro}. In general, $T_{\rm d, SED}$ does not necessarily provide a statistically sound representation of the PDF. For a discussion see Appendix \ref{appendix}.} grey body function. 

At $z\ge 5$ most of the sources observed with ALMA, when detected in dust continuum, have only a single (or very few) data-point at FIR wavelengths \citep[e.g.][]{Bouwens16,barisic2017dust,bowler2018obscured,big3drag,Tamura19}. Consequently, $T_{\rm d, SED}$ is assumed a priori in the fitting to reduce the degrees of freedom. The lack of knowledge of $T_{\rm d, SED}$ results in very large uncertainties on the derived galaxy properties, such as $M_{\rm d}$, the far-infrared luminosity $L_{\rm FIR}$, and obscured star formation rate \citep[SFR; for a detailed discussion see e.g.][]{nostro}.
Further observations in a larger number of ALMA bands would ameliorate the problem, but not necessarily solve it. Indeed, MIR wavelengths remain inaccessible to ALMA. Nevertheless, the inclusion of ALMA band $7-8-9$ data would improve the results significantly for galaxies at $z \simgt 5$. At these redshifts, these bands sample the SED at shorter wavelengths, closer to the FIR emission peak. 

Here we intend to overcome current observational limitations by combining dust continuum measurements with the widely observed fine-structure transition of singly ionized carbon \CII\ at $158\ \mathrm{\mu m}$. This line is the dominant coolant of the neutral atomic gas in the ISM \citep{Wolfire_2003}, making it one of the brightest FIR lines in most galaxies \citep{1991ApJ...373..423S}. Moreover, \CII\ has been proved to be connected to the SFR of local \citep{delooze14,Herrera_Camus_2015} and high-$z$ galaxies \citep[see e.g.][]{capak15,2015MNRAS.452...54M,Pentericci_2016,carniani:2017oiii,matthee2017alma,carniani2018clumps,carniani2018alma,2018ApJ...859...84H,2018smit,Carniani20}. 

In this work we propose a novel method for the dust temperature computation using $L_{\rm CII}$ as a proxy for the total gas mass, and therefore for $M_{\rm d}$ given a dust-to-gas ratio. Our method breaks the degeneracy between $M_{\rm d}$ and $T_{\rm d, SED}$ in the SED fitting. This allows us to constrain $T_{\rm d, SED}$ with a single continuum data point. As a byproduct of our method, we provide an interpretation of the empirical relation found by \citet{zanita19} between $M_{\rm H_{2}}$ and $L_{\rm CII}$. We also derive a more general relation connecting the \textit{total} gas mass $M_{\rm gas}$ with $L_{\rm CII}$. Joining the two we can also study the evolution of the molecular gas fraction $M_{\rm gas}/M_{\rm H_2}$ with redshift. 

The paper\footnote{Throughout the paper, we assume a flat Universe with the following cosmological parameters:  $\Omega_{\rm M}h^2 = 0.1428$, $\Omega_{\Lambda} = 1- \Omega_{\rm M}$, and $\Omega_{\rm B}h^2 = 0.02233$,  $h=67.32$, $\sigma_8=0.8101$, where $\Omega_{\rm M}$, $\Omega_{\Lambda}$, $\Omega_{\rm B}$ are the total matter, vacuum, and baryonic densities, in units of the critical density; $h$ is the Hubble constant in units of $100\, {\rm km s}^{-1}$, and $\sigma_8$ is the late-time fluctuation amplitude parameter \citep{Planck18}.} is organised as follows. We present our method for the dust temperature derivation in Sec. \ref{model}, and we test it on a sample of local galaxies (Sec. \ref{localtesting}). We then apply the method to the few high-$z$ galaxies ($z>4$, Sec. \ref{hztemplate}) for which multiple FIR continuum observations are available in the literature. In Sec. \ref{alfamolsec} we discuss additional outputs, i.e. the physical explanation for the relation by \cite{zanita19}, and the molecular gas fraction evolution with $z$. In Sec. \ref{summary} we summarise our results and discuss future applications.

\section{Method}\label{model}
Before introducing our method, we discuss two key ingredients, i.e. the dust-to-gas ratio $D$ and the conversion factor $\alpha_{\rm CII}= M_{\rm gas} / L_{\rm CII}$. Multiplying $L_{\rm CII}$ by the product $D \cdot \alpha_{\rm CII}$ we infer $M_{\rm d}$. We can then constrain $T_{\rm d, SED}$ using a single continuum data point. 

\subsection{Dust-to-gas ratio}\label{dtog}
Several studies \citep[e.g.][]{james2002, 2007ApJ...657..810D,galliano2008,Leroy_2011} have shown that $D$ scales linearly with metallicity, with little scatter, down to $Z \simlt 0.1\ Z_{\odot}$:
\begin{equation}\label{DtoZ}
    D=D_{\rm MW}\left(\frac{Z}{Z_{\odot}}\right),
\end{equation}
where $D_{\rm MW}=1/162$ is the Milky Way dust-to-gas ratio \citep{RemiR}. We adopt eq. \ref{DtoZ} as our fiducial choice, since almost all the galaxies to which we apply our method have metallicities $\simgt 0.2\ Z_{\odot}$. Hence, they are mostly unaffected by deviations\footnote{At very low metallicities ($Z \simlt 0.1\ Z_{\odot}$), deviations from the linear relation have been suggested \citep[see e.g.][]{2005galliano,2011A&A...532A..56G,RemiR,2019Devis}. For instance, \cite{RemiR} find a steeper $D-Z$ relation in their sample of local galaxies. 
However, the deviation is driven especially by the fewer, widely scattered data at $Z \leq 0.1\ Z_{\odot}$.}  from this linear scaling that might occur at $Z \simlt 0.1\ Z_{\odot}$. 

Moreover, the ideal targets of our method are the galaxies observed in current high-$z$ ALMA surveys (such as e.g. ALPINE, PI: Le F\'evre, and REBELS, PI: Bouwens), which are massive (stellar mass $M_{\star} \simeq 10^{10}\ M_\odot$), dusty, and evolved sources. From numerical simulations galaxies at $z\sim 6$ with similar stellar masses ($10^9<M_{\star}<10^{11}$) are expected to have $Z \simgt 0.1\ Z_\odot$ \citep{Ma16, Torrey19}. This is also confirmed, albeit with considerable uncertainties (relative errors up to $\sim 80\%$), by several studies which analyse FIR lines (such as \NII, \NIII, \CII, \CIII\ and \OIII) observations at $z \simgt 6-8$ to derive $Z$ \citep[see e.g.][and references therein]{ sostpereira, big3drag,2019A&A...631A.167D,Tamura19,Vallini20,Bakx20,2020arXiv200602447J,OIIImetal}. 

Current estimates of $Z$ at high redshift will be significantly ameliorated thanks to forthcoming ALMA observations and to the James Web Space Telescope (JWST) spectroscopy. Indeed, JWST will detect several optical nebular lines (such as \Hb, \Ha, \NII, \OII\ and \OIII) out to $z\sim 10$. This will allow us to reduce the relative errors associated to $Z$ down to $\sim 35\%$ even at very high-$z$ \citep[see e.g.][]{wright2010tracing,Maiolino_2019,Chevallard}, improving also our knowledge of the dust-to-gas ratios.

\subsection{\CII-to-total gas mass conversion factor}\label{alfaCIISFR}
The \CII\ conversion factor, $\alpha_{\rm CII}$, expresses the specific \CII\ emission efficiency per unit \textit{total} (i.e. atomic + molecular) gas mass. To investigate the relation between total $L_{\rm CII}$ and $M_{\rm gas}$, we use the following empirical relations\footnote{We adopt the standard units used for these quantities: surface star formation $[\mathrm{M_{\odot}\, kpc^{-2}\,yr^{-1}}]$, \CII\, luminosity $[\mathrm{L_{\odot}\, kpc^{-2}}]$, and gas density $[\mathrm{M_{\odot}\,kpc^{-2}}]$}:
\begin{align}
&\Sigma_{\rm SFR} = 10^{-6.99}\ \Sigma_{\rm CII}^{0.93} &\, \quad \rm{(De\  Looze\, relation)}\label{dLR}\\
&\Sigma_{\rm SFR} = 10^{-12}\ \kappa_s\ \Sigma_{\rm gas}^{1.4}& \, \quad \rm{(Kennicutt-Schmidt\, relation)}\label{KSR}\\
&\Sigma_{\rm gas}   = \alpha_{\rm CII} \Sigma_{\rm CII}& \, \quad \rm{(conversion\, relation)} \label{CVR}
\end{align}
The first relation has been inferred by \citet[][hereafter, DL]{delooze14} from the Dwarf Galaxy Survey (DGS) sample of local galaxies\footnote{For details on the DGS sample see Sec. \ref{localtesting}}. The second one is the Kennicutt–Schmidt relation \citep[][hereafter, KS]{kennicutt1998}. The \quotes{burstiness parameter} $\kappa_s$ quantifies the single sources deviations \citep[upwards for starbursts, and downwards for quiescent galaxies, see ][]{Heiderman_2010,ferraraCII} from the average relation. Finally, eq. \ref{CVR} is equivalent to the definition $\alpha_{\rm CII}=M_{\rm gas}/L_{\rm CII}$ under the assumption that \CII\ is spatially extended as the gas. 

We combine eq. \ref{dLR}-\ref{CVR} into the following one,
\begin{equation}\label{alfaKS}
    \alpha_{\rm CII}= \frac{11.3}{\kappa_s^{5/7}}\ \Sigma_{\rm SFR}^{-0.36} \quad \frac{M_\odot}{L_\odot}.
\end{equation}
This relation shows that satisfying the DL and KS relations at the same time implies that $\alpha_{\rm CII}$ cannot be constant. It must depend on the SFR and its mode (burst vs. quiescent). At a fixed SFR, galaxies with large $\kappa_s$ values (starbursts) have a lower $\alpha_{\rm CII}$ and therefore can produce a larger \CII\, luminosity per unit gas mass. The same is true if $\kappa_s$ is fixed and the SFR is larger. In high star formation regimes the more efficient \CII\, emission might depend on a more intense radiation field or higher gas density \citep{ferraraCII, pallottini:2019}. 

\subsubsection{Modification at high-$z$}

As we approach the Epoch of Reionization (EoR) a precise assessment of the KS relation becomes very difficult. \HI is not observable at $z \ge 4$, and typical $\mathrm{H_2}$ tracers (CO and dust) suffer from severe limitations\footnote{Observing CO transitions becomes challenging due to the larger cosmological distance of sources, and lower contrast against the Cosmic Microwave Background \citep[CMB, see e.g.][]{da2013effect}. This also makes dust emission observations more difficult at high-$z$. This is particularly true in the presence of cold dust nearly in equilibrium with the CMB \citep{da2013effect}. Most importantly, the impossibility to simultaneously constrain $M_{\rm d}$ and $T_{\rm d, SED}$ due to the few available data points, results in very large uncertainties on $M_{\rm d}$, and therefore $M_{\rm H_{2}}$.}. Hence $\Sigma_{\rm gas}$ is not reliably measurable. So far there is considerable evidence that FIR-detected galaxies at $z>5$ are strong UV emitters\footnote{This might, however, be due to an observational bias. Indeed, most high-z ALMA targets have been selected from UV observations (i.e. by construction they are strong UV emitters). There are few exceptions represented by the (sub)mm-selected targets, as in the surveys ASPECS \citep{2016ApJ...833...67W} and SPT \citep{weiss2013alma}. } with large SFRs, i.e. they are most likely starbursts ($\kappa_s \gg 1$, see e.g. \citealt{Vallini20},Vallini in prep.). 

The validity of the DL relation might also be questioned at high-$z$. Most studies agree that this relation is still valid at $z>4$, although its scatter is $\sim 2$ times larger than the local one \citep{carniani2018alma,carniani2018clumps,Matthee_2019,Schaerer_alpine}. However, in extreme cases (SFR$<30-50\ \mathrm{M_{\odot}/yr}$ or $z>8$) high-$z$ sources have been found to deviate more than $2 \sigma$ from the local DL relation, being systematically below the latter \citep{Pentericci_2016,knudsen2016merger,2017ApJ...836L...2B,Matthee_2019,laporte19}. 

Recently, \cite{Carniani20} showed that EoR galaxies lay on the slightly different (w.r.t. the one in eq. \ref{dLR}) DL relation appropriate for starburst/\HII-like galaxies\footnote{Which is also provided in \citealt{delooze14}}:
\begin{equation}\label{dlSFR}
    \Sigma_{\rm SFR}=10^{-7.06}\  y^2\  \Sigma_{\rm CII}  \, \quad \rm{(De Looze\, relation/starbursts).}
\end{equation}
once that obscured fraction of the SFR is appropriately included in $\Sigma_{\rm SFR}$. The factor $y=r_{\rm CII}/r_{\star}$ is introduced since there is growing evidence that at $z>4$ \CII\, emission is more extended than UV emission \citep[ $1.5 \simlt y \simlt 3$ at $z>4$, see e.g.][]{ carniani:2017oiii,carniani2018clumps,matthee2017alma,Matthee_2019,2019ApJ...887..107F,2020arXiv200300013F,2020A&A...633A..90G,Carniani20}. 
The origin of a such extended \CII\ structure is still debated. Current explanations range from emission by a) outflow remnants in the Circum Galactic Medium \citep[CGM, see e.g. ][]{Maiolino15,vallini:2015,10.1093/mnras/stx2458,2019ApJ...887..107F,10.1093/mnras/staa1163,2020A&A...633A..90G}, b) CGM gas illuminated by the galaxies strong radiation field \citep{carniani:2017oiii,carniani2018alma,2020arXiv200300013F}), to c) actively accreting satellites \citep{pallottini:17a,carniani2018clumps,Matthee_2019}.

By combining eq. \ref{dlSFR} with eqs. \ref{KSR} and \ref{CVR}, we derive the high-$z$ conversion factor 
\begin{equation}\label{alfahz}
    \alpha_{\rm CII,hz}=\frac{32.47}{\kappa_s^{5/7}}\ y^{2}\ \Sigma_{\rm SFR}^{-0.29} \quad \frac{M_\odot}{L_\odot}.
\end{equation}
Using the DL relation for starbusts, independently on the chosen factor $y$, results in a rescaling upwards of $\alpha_{\rm CII}$ at high-$z$ with respect to $z\simeq 0$. The dependence on $\Sigma_{\rm SFR}$ and $\kappa_s$ is almost unchanged. Additionally, at a fixed SFR and $\kappa_s$, galaxies with lower $y$ (less extended \CII\ emission) have a lower $\alpha_{\rm CII}$, i.e. a larger \CII\, luminosity per unit gas mass.

\subsection{DUST TEMPERATURE}\label{tdustcomput}

We assume an optically thin, single-temperature, grey-body approximation. 
The dust continuum flux $F_{\rm \nu}$ observed against the CMB
at rest-frame frequency $\nu$ can be written as \citep[see e.g.][]{da2013effect, Kohandel19}
 \begin{equation}\label{flux_eq}
F_{\rm \nu} =  g(z) M_{\rm d} \kappa_{\nu} [B_{\nu}(T_{\rm d,SED}')-B_{\nu}(T_{\rm CMB})],
\end{equation}
where $g(z) = {(1+z)}/{d_L^2}$, $d_{\rm L}$ is the luminosity distance to redshift $z$, $k_{\rm \nu}$ is the dust opacity, $B_\nu$ is the black-body spectrum, and $T_{\rm CMB}(z)$ is the CMB temperature\footnote{$T_{\rm CMB}(z)=T_{\rm CMB,0}(1+z)$, with $T_{\rm CMB,0}=2.7255\, \mathrm{K}$ \citep{Fixsen09}} at redshift $z$.

At wavelengths $\lambda > 20\,\mathrm{\mu m}$, $k_{\rm \nu}$ can be approximated as \citep{draine2004astrophysics}
\begin{equation}\label{opacity}
\kappa_{\nu} = \kappa_* \left(\frac{\nu}{\nu_{*}}\right)^{\beta}.
\end{equation} 
where the choice of $(\kappa_*, \nu_*, \beta)$ depends on the assumed dust properties. We consider Milky Way-like dust, for which standard values are $(k_*, \nu_*, \beta)$ = (52.2 ${\rm cm^2 g^{-1}}$, $2998\, {\rm GHz}$, 2), see \cite{10.1111/j.1365-2966.2009.16164.x}.
We also account for the fact that the CMB acts as a thermal bath for dust grains, setting a lower limit for their temperature. We correct $T_{\rm d, SED}$ for this effect, following the prescription\footnote{$T_{\rm d,SED}' = \{T_{\rm d,SED}^{4+\beta}+T_{\rm CMB,0}^{4+\beta}[(1+z)^{4+\beta}-1]\}^{1/(4+\beta)}$. In the following, we drop the apex from the dust temperature symbol for better readability. It is then intended we always refer to the \textit{CMB-corrected} dust temperature.} by \cite{da2013effect}.

Eq. \ref{flux_eq} has two parameters, $M_{\rm d}$ and $T_{\rm d, SED}$. \CII\ observations can be used to determine $M_d$:  
\begin{equation}\label{zanella}
M_d = D M_{\rm gas} = D\, \alpha_{\rm CII} L_{\rm CII}. 
\end{equation}
We substitute in eq. \ref{flux_eq} and specialize to the \CII\ line frequency $\nu_{0} =1900.54$ GHz. 
We thus introduce the \CII\,-based dust temperature $T_{\rm d, CII}$, defined as the solution of  \begin{equation}\label{flux_eq_sost}
F_{\nu_ 0} =  g(z) D\, \alpha_{\rm CII} L_{\rm CII}  \kappa_{\nu_0} [B_{\nu_0}(T_{\rm d,CII})-B_{\nu_0}(T_{\rm CMB})].
\end{equation}
We can re-write this equation in a more compact form, yielding the explicit expression for $T_{\rm d, CII}$:
\begin{equation}\label{td1pto}
T_{\rm d, CII} = \frac{T_{0}}{\ln(1+f^{-1})}.
\end{equation}
where $T_0= h_{\rm P} \nu_0/ k_{\rm B} = 91.86$ K is the temperature corresponding to the \CII\ transition energy ($k_{\rm B}$ and $h_{\rm P}$ are the Boltzmann and Planck constants). We have defined: 
\begin{equation}\label{f}
f = {\cal B}(T_{\rm CMB}) + A^{-1}\tilde F_{\nu_0},
\end{equation}
where ${\cal B}(T_{\rm CMB})= [\exp(T_0/T_{\rm CMB})-1)]^{-1}$. The non-dimensional continuum flux $\tilde F_{\nu_0}$ and the constant $A$ are defined as
\begin{equation}\label{defs}
\begin{split}
    &\tilde F_{\nu_0} = \frac{\lambda_0^2 F_{\nu_0}}{2 k_{\rm B} T_0} = 0.98 \times 10^{-16} \left(\frac{F_{\nu_0}}{\rm mJy} \right),\\
    & A =  g(z) \alpha_{\rm CII} D L_{\rm CII} k_0 = 5.2 \times 10^{-24} \left[\frac{g(z)}{g(6)}\right] \left(\frac{L_{\rm CII}}{L_\odot}\right)\left(\frac{\alpha_{\rm CII}}{M_\odot/L_\odot}\right)  D.  \\
\end{split}
\end{equation}
Clearly, if $\tilde F_{\nu_0}/ A \gg {\cal B}(T_{\rm CMB})$ the CMB effects on dust temperature become negligible.

Eq. \ref{td1pto} can be used to compute $T_{\rm CII}$ using a single $1900$ GHz observation (which provides both $L_{\rm CII}$ and $F_{ \nu_0}$) once one has an estimate for the two parameters $D$ (Sec. \ref{dtog}) and $\alpha_{\rm CII}$ (Sec. \ref{alfaCIISFR}). 

\subsubsection{Numerical implementation}\label{MC}
Writing explicitly the expressions for $D$ and $\alpha_{\rm CII}$ in eq. \ref{defs}, we can show that $T_{\rm d, CII}$ is ultimately a function of the following parameters $(\kappa_s,z,F_{\nu_ 0},Z,\Sigma_{\rm SFR},L_{\rm CII}) $. 
For local galaxies, all these quantities are well constrained by observations. In practice we solve eq. \ref{td1pto} performing a random sampling of these parameters around the measured values, within the uncertainties. 
Differently, at high-$z$ $\kappa_s$ is largely unknown\footnote{At high-$z$ we also introduce the parameter $y$. This is often well constrained by observations.}. Hence we consider a broad random uniform distribution for this parameter. 

To constrain $T_{\rm d, CII}$ at high-$z$, we add the following physical conditions: 
\begin{enumerate}
    \item $M_{\rm d}$ does not exceed the largest dust mass producible by supernovae (SNe), $M_{\rm d, max}$. To quantify $M_{\rm d, max}$ we take a metal yield constraint $y_Z <2\ \mathrm{M_{\odot}}$ per SN, and assume that all the produced metals are later included in dust grains. Then:
    \begin{equation}\label{mdmax}
    M_{\rm d, max}= y_Z \nu_{\rm SN} M_{\star}
    \end{equation}
    where $\nu_{\rm SN} = (53\ \mathrm{M_{\odot}})^{-1}$ is the number of SNe per solar mass of stars formed for a standard Salpeter 1-100 $M_\odot$ IMF \citep{10.1046/j.1365-8711.2000.03209.x}.
    \item $\mathrm{SFR_{\rm FIR}} \sim 10^{-10}\ L_{\rm FIR}$ \citep{kennicutt1998}, does not exceed the total measured SFR. This directly relates to the dust mass and temperature as $L_{\rm FIR}=M_{\rm d}\ (T_{\rm d, CII}/6.73)^6$ \citep{10.1111/j.1365-2966.2009.16164.x}.
\end{enumerate}

$T_{\rm d, CII}$ solutions not satisfying (i) and (ii) are discarded. These conditions result in a lower (upper) cut for very cold (hot) dust temperatures corresponding to unphysically large dust masses (FIR luminosity and SFR). This allows us to effectively constrain $T_{\rm d, CII}$ at high-$z$ despite the lack of information on $\kappa_s$.

\begin{table*}
  \begin{center}
    \begin{tabular}{l|c|c|c|c|c|c|c||c|c|c|r}
        \hline
         ID & Galaxy  & $\nu_{2}$         & $F_{\nu_2}$     & $d_{\rm L}$     & $F_{\nu_0}$       & $Z$           & $ \log \Sigma_{\rm SFR}$  & $\log L_{\rm CII}$ & $\kappa_s$ & $T_{\rm d, SED}$ & $T_{\rm d,CII}$\\
            &     & $ [\mathrm{GHz}]$ & $[\mathrm{Jy}]$     &$[\mathrm{Mpc}]$ & $[\mathrm{Jy}]$   & $[Z_{\odot}]$ & $\mathrm{[M_{\odot}\ yr^{-1}\ kpc^{-2}]}$ & $\mathrm{[L_{\odot}]}$ & & $\mathrm{[K]}$ & $\mathrm{[K]}$\\
        \hline
    0&   UGC4483$^{a}$ & $4285$       & $0.109\pm0.007$       & $3.2$  & $0.081\pm0.029$      &  $0.053$ & $-3.041$           & $4.119$ & $0.1$ & $ 31.0 ^{+ 5.0 }_{- 2.0 }$ & $ 32.0 ^{+ 7.0 }_{- 7.0 }$\\
       
    1&   VIIZw403$^{a}$ & $4285$      &	$0.493\pm0.026$       & $4.5$  & $0.260\pm0.037$        & $0.083$  & $-2.621$           & $4.994$ & $0.3$ &  $ 34.0 ^{+ 2.0 }_{- 1.0 }$ & $ 30.0 ^{+ 4.0 }_{- 6.0 }$\\

    2&   NGC 1569$^{a}$ & $4285$      &$42.600\pm 2.100$      & $3.1$  & $39.700\pm 4.800$       & $0.190$  & $-1.721$           & $6.669$ & $3.1$ & $ 29.0 ^{+ 1.0 }_{- 1.0 }$ & $ 34.0 ^{+ 5.0 }_{- 7.0 }$\\

    3&   II Zw 40$^{a}$ & $4285$      &$5.580\pm 0.280$       & $12.1$ & $3.140\pm 0.431$          & $0.309$ & $-0.818$           & $6.586$ & $0.9$ &  $ 33.0 ^{+ 2.0 }_{- 1.0 }$ & $ 43.0 ^{+ 7.0 }_{- 11.0 }$\\	

    4&   NGC 4214$^{a}$ & $4285$      &$20.400\pm 1.020$      & $2.9$  & $23.400\pm 2.820$             & $0.331$  & $-3.054$          & $5.977$& $0.4$ &  $ 28.0 ^{+ 1.0 }_{- 1.0 }$ & $ 27.0 ^{+ 3.0 }_{- 5.0 }$\\

    5&   UM 448$^{a}$ & $4285$        &$4.040\pm 0.203$       & $87.8$ & $2.850\pm 0.345$         & $0.380$  & $-0.957$          & $8.281$ & $0.8$ & $ 31.0 ^{+ 1.0 }_{- 1.0 }$ & $ 37.0 ^{+ 6.0 }_{- 8.0 }$\\	

    6&   NGC1140$^{a}$ & $4285$       &$3.430\pm 0.172$      & $20$    & $4.050\pm 0.487$             & $0.436$  & $-2.096$           & $7.169 $ & $0.1$ & $ 27.0 ^{+ 1.0 }_{- 1.0 }$ & $ 26.0 ^{+ 3.0 }_{- 5.0 }$\\	
       \hline
       
    7&   LARS2$^{b}$ & $4285$         &$0.080\pm0.011$       & $131.4$  & $0.121\pm0.014$            & $0.309$   & $-2.901$               & $7.230$ & $0.1$ &   $ 26.0 ^{+ 1.0 }_{- 1.0 }$ & $ 24.0 ^{+ 7.0 }_{- 5.0 }$\\
       
    8&   LARS3$^{b}$ & $4285$         &$9.306\pm0.008$	     & $138.7$  &$4.745\pm0.019$              & $0.468$   & $-1.444$              & $9.190$ & $2.7$ &  $ 34.0 ^{+ 1.0 }_{- 1.0 }$ & $ 24.0 ^{+ 4.0 }_{- 4.0 }$\\

    9&   LARS8$^{b}$ & $2998$         &$4.322\pm0.027$       & $169.6$  &$3.346\pm0.034$             & $0.589$   & $-1.850$              & $9.550$ & $1.1$ & $ 24.0 ^{+ 2.0 }_{- 2.0 }$ & $ 18.0 ^{+ 2.0 }_{- 2.0 }$\\
       
    10&   LARS9$^{b}$ & $4285$         &$1.147\pm0.017$       & $208.7$  &$1.306\pm0.027$              & $0.427$   & $-0.425$              & $9.170$ & $0.3$ &  $ 25.0 ^{+ 2.0 }_{- 2.0 }$ & $ 28.0 ^{+ 5.0 }_{- 6.0 }$\\
       
    11&   LARS12$^{b}$ & $4285$        &$0.104\pm0.004$       & $473.8$  &$0.062\pm0.004$             & $0.398$    & $-2.021$              & $8.580$ & $5.1$ &  $ 33.0 ^{+ 1.0 }_{- 1.0 }$ & $ 20.0 ^{+ 3.0 }_{- 3.0 }$\\
       
    12&   LARS13$^{b}$ & $4283$        &$0.506\pm0.003$       & $701.1$  &$0.286\pm0.003$             & $0.575$   & $-1.333$              & $9.250$ & $4.5$ & $ 33.0 ^{+ 1.0 }_{- 1.0 }$ & $ 25.0 ^{+ 4.0 }_{- 5.0 }$\\
       \hline

    13&   NGC4631$^{c}$ & $3409$       &$31.11$     & $6.2$ &$34.23$        & $0.50$    & $-3.18$    & $6.91$ & $0.4$ &  $ 24.0 ^{+ 3.0 }_{- 2.0 }$ & $ 22.0 ^{+ 3.0 }_{- 3.0 }$\\

    14&   NGC3627$^{c}$ & $3409$       &$20.68$     & $9.3$ &$17.40$         & $1.77$   & $-2.85$   & $6.46$ & $0.7$ &  $ 27.0 ^{+ 3.0 }_{- 3.0 }$ & $ 23.0 ^{+ 3.0 }_{- 4.0 }$\\

    15&   NGC2146$^{c}$ & $3409$       &$137.93$    & $12.9$ &$229.17$      & $0.98$   & $-2.03$    & $8.28$ & $0.1$ &  $ 21.0 ^{+ 2.0 }_{- 2.0 }$ & $ 25.0 ^{+ 4.0 }_{- 5.0 }$\\       

    16&   NGC3938$^{c}$ & $3409$       &$30.41$     & $17.9$  &$38.72$     & $2.10$    & $-2.98$     & $6.80$ & $0.1$ & $ 23.0 ^{+ 2.0 }_{- 2.0 }$ & $ 31.0 ^{+ 6.0 }_{- 7.0 }$\\ 

    17&   M83$^{c}$ & $3409$           &$135.78$    & $7.4$  &$111.93$       & $1.23$   & $-2.87$    & $7.37$ & $0.5$ & $ 27.0 ^{+ 4.0 }_{- 3.0 }$ & $ 21.0 ^{+ 3.0 }_{- 3.0 }$\\           
       
    18&   M82$^{c}$ & $3409$           &$1117.00$   & $2.9$  &$614.57$     & $2.75$  & $-1.56$     & $7.38$ & $5.9$ &   $ 32.0 ^{+ 5.0 }_{- 4.0 }$ & $ 21.0 ^{+ 3.0 }_{- 3.0 }$\\           
      \hline

    \end{tabular}
    \caption{Properties of galaxies included in our benchmark local sample. For the data without specified uncertainty, we consider a $20\%$ relative error as a conservative choice. \textbf{References}:$^{a}$\citep{radiiDL14sample,delooze14,continuumDGS,CIIDGS}, $^{b}$\citep{stefanosample}, and $^{c}$\citep{fernandez2016vizier,leroy2008SFR,SFRmessicani,simonaRADII,NGC3938}. }
     \label{tabloc} 
  \end{center}
\end{table*}

\section{Local testing}\label{localtesting}
We have selected $19$ local galaxies for which the needed data are available: (a) $\kappa_s$, (b) redshift, (c) metallicity, (d) total SFR and $\Sigma_{\rm SFR}$, (e) total $L_{\rm CII}$, (f) at least two FIR continuum detections, one of which at $\nu_0$. These galaxies are drawn from the following catalogs\footnote{Other local samples, such as KINGFISH \citep[][]{kingfish} and GOALS surveys \citep[][]{2017ApJS..229...25C}, lack one of the required data (total \CII\ luminosity and metallicity, respectively).}: 
\begin{itemize}
    \item Dwarf Galaxy Survey \citep[DGS, see e.g.][]{delooze14,2014PASP..126.1079M,2020arXiv200900649M}: targeting a total of $50$ local dwarf galaxies, whose \CII, \OI\ and  \OIII\ line emission are mapped with the Hershel Space Observatory; 
    
    \item Lyman Alpha Reference Sample \citep[LARS, see e.g.][]{Hayes_2014,Ostlin_2014}: consisting of 14 low-redshift ($z= 0.03 - 0.2$) mildly starbursting systems observed in multiple bands with HST. This sample was intended as a local laboratory for the study of Ly$\alpha$, which is one of the dominant lines used to characterise high-$z$ sources; 
    
    \item The complete database of the Hershel/Photoconductor Array Camera and Spectrometer \citep[PACS, see][]{fernandez2016vizier}: a coherent database of spectroscopic observations of FIR fine-structure lines (in the range $10 - 600\ \mathrm{\mu m}$) collected from the Herschel/PACS spectrometer archive for a local sample of $170$ Active Galactic Nuclei (AGNs), $20$ starburst, and $43$ dwarf galaxies. 
    
\end{itemize}
The selected galaxies and their properties are reported in Tab. \ref{tabloc}. Hereafter we refer to these galaxies as the \textit{local sample}.

\begin{figure*}
    \centering
    \includegraphics[width=0.493\linewidth]{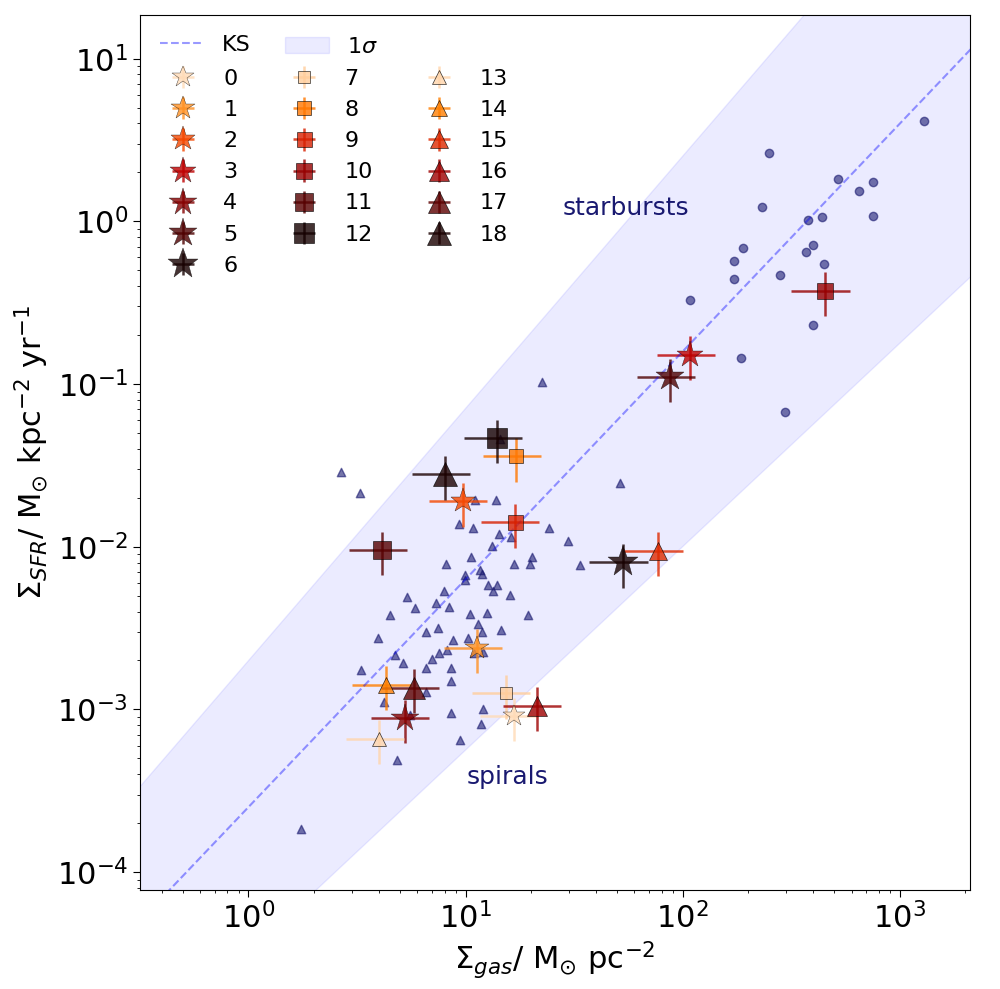}
    \includegraphics[width=0.49\linewidth]{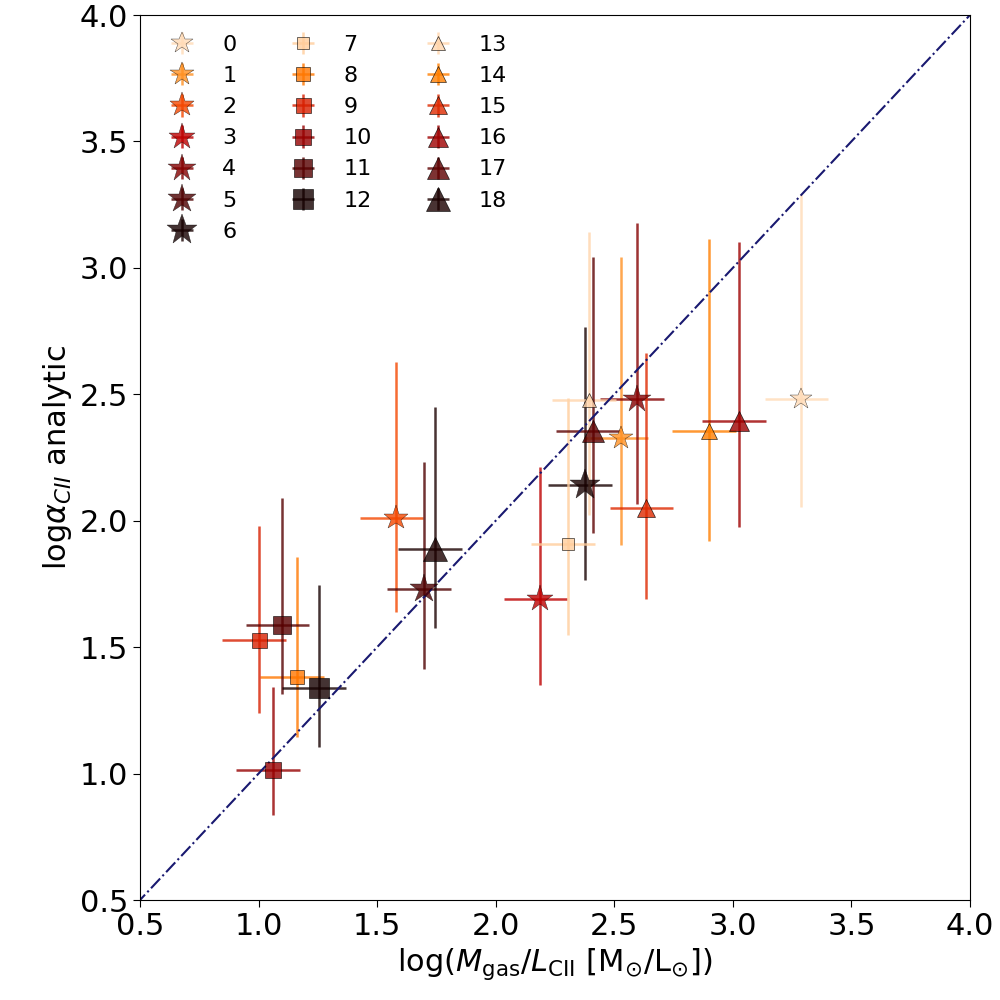}
    \caption{\textbf{Left panel}: Measured $\Sigma_{\rm SFR}$ vs. $\Sigma_{\rm gas}$ of our local sample. We associate to each galaxy an ID number which will be used in the following plots. The dashed blue line represents the KS relation (eq. \ref{KSR}) and the blue shaded region its intrinsic scatter. Also shown for reference are a number of local spirals (black triangles, \citealt{spirstarb}) and starbursts (black stars, \citealt{spirstarb}). We distinguish each galaxy in our local sample with a different colour and identify them in the legend with their IDs as in Tab. \ref{tabloc}. We also differentiate the three sub-samples with a different shape: a (star), b (square), and c (triangle, all references are the same as in Tab. \ref{tabloc}). We note galaxies in the local sample are consistent within errors with the KS relation. 
    \textbf{Right panel}: \CII\ conversion factor computed from eq. \ref{alfaKS} vs. the observed $\log (M_{\rm gas}/L_{\rm CII})$ for the same galaxies as in the left panel. The solid symbols correspond to the value of $\alpha_{\rm CII}$ obtained considering for each galaxy the $\kappa_s$ value computed from the measured $\Sigma_{\rm SFR}$ and $\Sigma_{\rm gas}$. 
    The dotted dashed black line is the bisector, i.e. it represents the relation $\log \alpha_{\rm CII} = \log (M_{\rm gas}/L_{\rm CII})$. The fact that the points are lay within $\simlt 1.5 \sigma$ from the bisector shows that eq. \ref{alfaCIISFR} gives a good estimate of the observed gas mass-to-\CII\ luminosity ratio. }
    \label{KS}

    \includegraphics[width=0.55\linewidth]{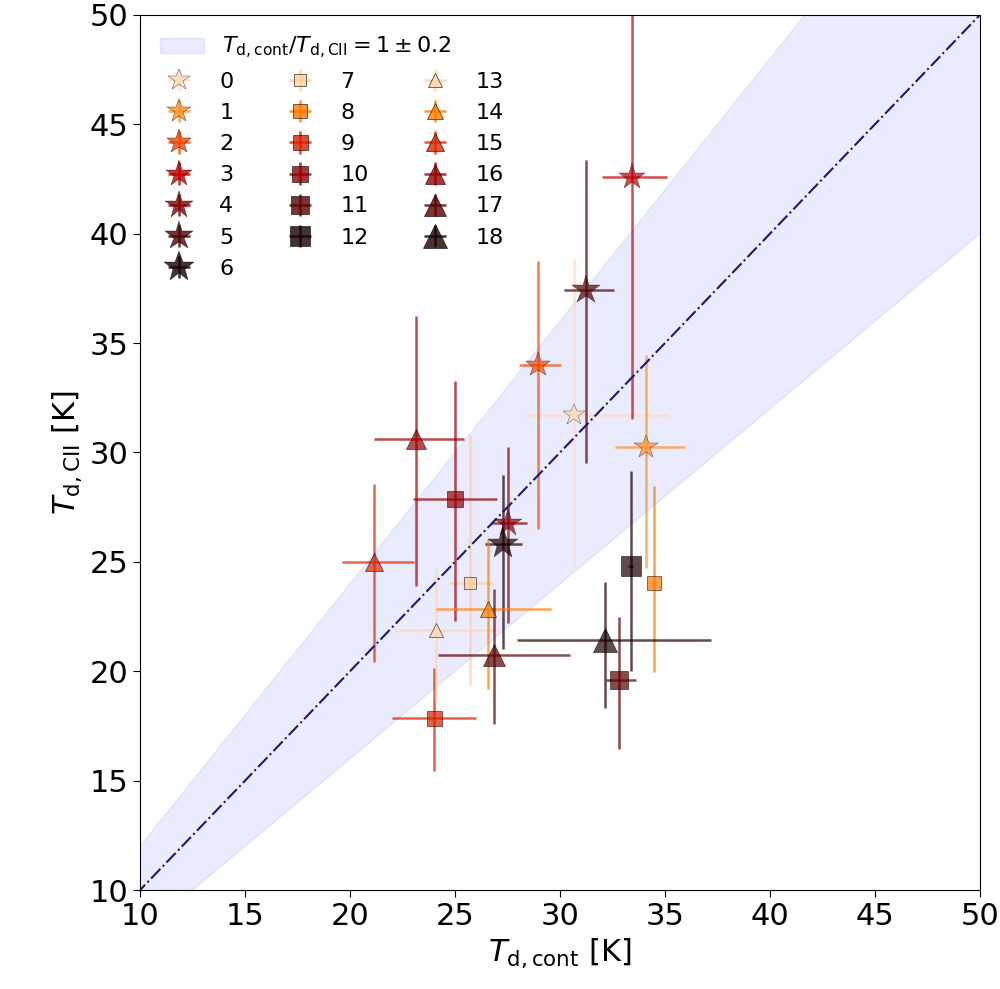}
    \caption{Comparison between $T_{\rm d,CII}$ (from eq. \ref{td1pto}) and $T_{\rm d, SED}$ (from eq. \ref{flratio}) in our local template sample of galaxies (see Tab. \ref{tabloc} for the properties of each galaxy corresponding to the ID in legend). The dotted dashed grey line represents the relation $T_{\rm d,CII}=T_{\rm d, SED}$ and the shaded area a deviation from the equality of $\pm 20\%$. }
    \label{Td1vsTd2}
\end{figure*}

The DL relation (eq. \ref{dLR}) has been derived from a portion of this same sample and therefore is nearly satisfied by construction. The galaxies in the local sample also follow the KS relation with a scatter consistent with that of local spirals and starbursts ($0.1\le \kappa_s\le 5.9$, see Fig. \ref{KS}, left panel). 

We compare the value of $\alpha_{\rm CII}$ resulting from eq. \ref{alfaKS} with the ratio $M_{\rm gas}/L_{\rm CII}$ derived from observations (Fig. \ref{KS}, right panel). 
We find $0.7\le \log \alpha_{\rm CII} \le 3.2$. The predicted $\alpha_{\rm CII}$ are consistent with the data at $< 1.5 \sigma$, although there are significant uncertainties. 

Finally, we compare $T_{\rm d, CII}$ and $T_{\rm d,SED}$. For the local sample galaxies we deduce $T_{\rm d,SED}$ from the following equation:
\begin{equation}\label{flratio}
     \frac{F_{\nu_{1}}}{F_{\nu_{2}}}= \frac{\kappa_{\nu_{1}} [B_{\nu_{1}}(T_{\rm d, SED})-B_{\nu_{1}}(T_{\rm CMB})]}{\kappa_{\nu_{2}} [B_{\nu_{2}}(T_{\rm d, SED})-B_{\nu_{2}}(T_{\rm CMB})]}.
\end{equation}
where we consider\footnote{We select these frequencies to avoid PAH contamination present at wavelengths $ \simlt 20\ \mathrm{\mu m}$. For ID10 we take $\nu_{\rm 2}=2998$ ($100\ \mathrm{\mu m}$) since observations at the above mentioned frequencies are not available.} $\nu_{1}=1900$ GHz and $\nu_{2}=(4285,3409)$ GHz, corresponding to $\lambda_{2}=(70,88)\ \mathrm{\mu m}$. This method based on the continuum fluxes ratio is equivalent to the single temperature grey-body SED fitting\footnote{For all the sources where multiple continuum observations are available we also computed the full SED fitting. We find $T_{\rm d, SED}$ values fully consistent with that obtained from continuum fluxes ratio.}, hence the obtained dust temperature is indeed $T_{\rm d, SED}$. 

To produce $T_{\rm d,CII}$ we consider a flat distribution for the burstiness parameter $0.1 \simlt \kappa_{\rm s} \simlt 5.9$, where the range is derived from gas masses and SFR measurements for the single sources (see Tab. \ref{tabloc}). The temperatures comparison is shown in Fig. \ref{Td1vsTd2}. We find that $T_{\rm d,CII}$ and $T_{\rm d,SED}$ are consistent within a $20\%$ uncertainty in $\sim 90\%$ of the sources (precisely, all but galaxies ID11, ID18). The two discrepant sources are the most bursty ones in the sample, with an inferred $\kappa_{\rm s} \sim 5$. Considering $\kappa_{\rm s} \sim 5$ in these two cases (rather than the aforementioned flat distribution) would allow us to correctly recover $T_{\rm d, SED}$ within $20\%$ uncertainty. Nevertheless, we have preferred to consider a flat distribution for $\kappa_{\rm s}$ to be more consistent with a general application at high-$z$, where this parameter is almost always unconstrained.

\begin{table*}
  \begin{center}
    \begin{tabular}{l|c|c|c|c|c|c|c|r}
        \hline
        Galaxy & $z$ & $F_{\nu_0}$ & $Z$ & $ \log \Sigma_{\rm SFR}$  & $ L_{\rm CII}$ & $\kappa_{\rm s}$& $y=r_{\rm CII}/r_{\rm *}$ & $M_{\star}$\\
                &   & $[\mathrm{mJy}]$ & $[Z_{\odot}]$ & $\mathrm{[M_{\odot}\ yr^{-1}\ kpc^{-2}]}$ & $[10^{8}\ \mathrm{L_{\odot}]}$& & & $[10^{8}\ \mathrm{M_{\odot}}]$ \\
        \hline\hline
     SPT0418-47$^{a}$  &  $4.23$ & $1.38\pm0.25$         & $0.30-1.30$    &  $1.72$ & $19.9\pm 1.5$ & $9$ & $1.5$ & $120.0\pm 15.0$ \\
     
     B1465666$^{b}$    & $7.15$   & $0.13\pm 0.03$        & $0.40\pm0.30$  & $1.32$  &  $11.0\pm 1.4$ & $-$ & $1.4\pm 0.4$ & $7.7$ \\
     
     MACS0416-Y1$^{c}$ & $8.31$   & $<0.02$                & $0.20\pm0.16$  & $1.16$  &  $1.4\pm 0.2$ & $-$  & $1.2 \pm 0.4$ & $2.0$\\
          
      \hline

    \end{tabular}
    \caption{Properties of our high-$z$ template sample of galaxies. We underline that for the data where the uncertainty is not given we consider a $30\%$ relative error which is a conservative choice given the other available data. \textbf{References}:$^{a}$\citet{2017BothwellSPT,2019A&A...631A.167D,ReuterSPT,2020Rizzo}. Here we show the intrinsic values, which are obtained by dividing by the magnification factor of the source $\mu = 32.7$ \citep{2019A&A...631A.167D}; $^{b}$\citet{big3drag}, and $^{c}$\citet{Bakx20}.}
     \label{tabhz} 
  \end{center}
\end{table*}
\section{Application at high redshift}\label{hztemplate} 
 We now apply our method to high-$z$ galaxies.  
We have collected a small (3 galaxies) sample for which the properties b)-f) are measured. 
The high-$z$ sample contains:
\begin{itemize}
    \item SPT 0418-47 \citep{weiss2013alma,strandet2016redshift}: a strongly lensed Dusty Star-Forming Galaxy (DSFG) at redshift $z=4.225$; 
    
    \item  B1465666 \citep[i.e. the \quotes{big three dragons}, ][]{big3drag}: a Lyman Break Galaxy (LBG) at $z=7.15$ 
    
    \item MACS0416-Y1 \citep{Tamura19,Bakx20}: an LBG at $z\sim 8.31$ 
\end{itemize}
The  properties of these galaxies are summarized in Tab. \ref{tabhz}. These sources are UV-selected, highly star forming (yet not extreme as SFR$ \simlt 500\ \mathrm{M_{\odot}/yr}$), do not host AGN, and are presumably main sequence high-$z$ galaxies.

For all these sources the parameter $y=r_{\rm CII}/r_{\star}$ has been estimated \citep{2020Rizzo,big3drag,Bakx20}. Moreover, \cite{2020Rizzo} provided a constraint on $\Sigma_{\rm gas}$ for SPT 0418-47. Hence, we can also derive the burstiness parameter for this galaxy finding $\kappa_{\rm s} \sim 9$. Due to the uncertainty in the gas mass derivation, in the computation we consider a Gaussian distribution centred around this value with a $\sigma \sim 2$ (i.e. we allow for values in the range $3 \simlt \kappa_{\rm s} \simlt 15$, consistent with previous works e.g. \citealt{Vallini20}).

For the remaining two galaxies $\kappa_{\rm s}$ is unknown, hence we need to define a broader distribution of possible values for this parameter. High redshift UV selected sources are strong UV emitters by construction, highly star forming, and consequently they are expected to be starburst $\kappa_s > 1$. Both locally and at intermediate redshift, values up to $\kappa_s \simeq 100$ have been observed in such galaxies \citep[see e.g.][]{Daddi_2010}. Recently, \cite{Vallini20} for the mildly star-bursting COS-3018 at $z=6.854$  found $\kappa_s\sim 3$, applying the \CII-emission model given in \cite{ferraraCII}. Applying the same method to B1465666 and MACS0416-Y1 Vallini in prep. finds very large values $30 \simlt k_{\rm s} \simlt 140$. Hence, we conservatively choose a random uniform distribution in the range $10<\kappa_s<100$ for these two sources.

Using eq. \ref{alfahz} we compute the coefficient $\alpha_{\rm CII, hz}$. We find, $\alpha_{\rm CII,hz}=5^{+2}_{-1}$ for SPT 0418-47, which is consistent with the recent estimate by \cite{2020Rizzo} of $M_{\rm gas}/L_{\rm CII} \sim \alpha_{\rm CII,hz}=7\pm 1$. We derive $\alpha_{\rm CII,hz}=2^{+2}_{-1}$ for B14-65666, and $\alpha_{\rm CII,hz}=2^{+2}_{-1}$  for MACS0416-Y1. We note that these values are lower than the average $\alpha_{\rm CII}$ values found locally ($z\sim 0$, see e.g. Fig. \ref{KS}). This might indicate a trend of less efficient \CII\ emission per unit gas mass at higher redshift. 

We now compare the $T_{\rm d, CII}$ estimated by our model with the $T_{\rm d, SED}$ from the SEDs. We summarise our findings and compare with literature data in Tab. \ref{tabhz1}. 
For SPT 0418-47 we derive $T_{\rm d, CII}=49^{+9}_{-8}\ \mathrm{K}$ that is consistent, within the error, with the $T_{\rm d, SED}$ from the SED fitting by \cite{strandet2016redshift} ($T_{\rm d, SED}=45\pm 2\ \mathrm{K}$). Recently \cite{ReuterSPT} derived a slightly higher dust temperature $T_{\rm d, SED}=58 \pm 11\ \mathrm{K}$ for this source, which is still consistent with our result\footnote{\cite{ReuterSPT} left $\lambda_{\star}=100\ \mathrm{\mu m}$ as an additional free parameter in the SED fitting \citep[see also][for a detailed discussion]{spilker2016alma}, which resulted in a larger ($\times 4$) uncertainty alongside a raise in the dust temperature. The fewer FIR data currently available at very high-$z$ do not allow for the application of a similar fitting procedure on a large scale. Hence, at the current stage a simpler grey-body (as in \citealt{strandet2016redshift}) with little variation in the dust properties is uniformly applied, leading to pretty consistent $T_{\rm d, SED}$ derivations for different sources.}.
Although the uncertainty of our $T_{\rm d, CII}$ is larger than the one of $T_{\rm d, SED}$ this is somewhat expected. The SED for SPT 0418-47 is well constrained, featuring data points on both sides of the FIR spectrum. On the other hand, the metallicity of the source is very uncertain, $0.3 \le Z/Z_{\odot} \le 1.3$ and this affects directly the error on our $T_{\rm d, CII}$.

For B14-65666 we find $T_{\rm d,CII}=61^{+16}_{-15}\ \mathrm{K}$. This value is consistent with $T_{\rm d, SED}=48-61\ \mathrm{K}$ which is inferred considering  $1.5<\beta < 3.0$ \citep{big3drag}. 
For MACS0416-Y1 \citep{Tamura19,Bakx20} we consider the upper limit on $F_{\nu_0}$ recently derived by \cite{Bakx20}. Interestingly $T_{\rm d, CII}$ in eq. \ref{td1pto} decreases with $f\propto F_{\rm \nu_0}$. Hence, for this galaxy we provide an upper limit for the \CII\ derived dust temperature: $T_{\rm d,CII}\le 82^{+16}_{-19}\ \mathrm{K}$. Very hot dust temperatures $T_{\rm d, CII}>120\ \mathrm{K}$ are excluded thanks to the condition on $\mathrm{SFR_{\rm FIR}}$ (see Sec. \ref{MC})\footnote{Without the condition on $\mathrm{SFR}_{\rm FIR}$ dust temperatures as large as $T_{\rm d, CII} \simgt 130\ \mathrm{K}$ would be reached. In part, this is a consequence of the very large uncertainty ($\sim 80\%$) on the already low metallicity of this galaxy ($Z=0.2\ \mathrm{Z_{\odot}}$). Indeed for a fixed flux $F_{\rm \nu_0}$, $T_{\rm d, CII}$ diverges as the metallicity $Z\rightarrow 0$ as this is equivalent to $M_{\rm d} \rightarrow 0$, see eq. \ref{zanella}.}.
This result is particularly relevant as \cite{Bakx20} obtained only a lower limit for the dust temperature $T_{\rm d, SED} \simgt 80\ \mathrm{K}$. By combining the two results we can constrain the dust temperature of MACS0416-Y1 in the range $T_{\rm d, SED} \sim 80-98\ \mathrm{K}$.

In conclusion, with our method, we can provide dust temperature estimations comparably accurate as that obtained from the traditional SED fitting with multiple bands data out to $z=8.31$. This is very encouraging, as for the single high-$z$ sources targeted by large programs, only single band measurements are generally available. Hence, commonly used SED fitting is not applicable without some underlying restrictive assumption on $T_{\rm d, SED}$. Our method can be used in these cases to improve the accuracy of the interpretation of FIR observations, and derive dust and galaxies properties.  

\begin{figure*}
    \centering
    \includegraphics[width=0.8\linewidth]{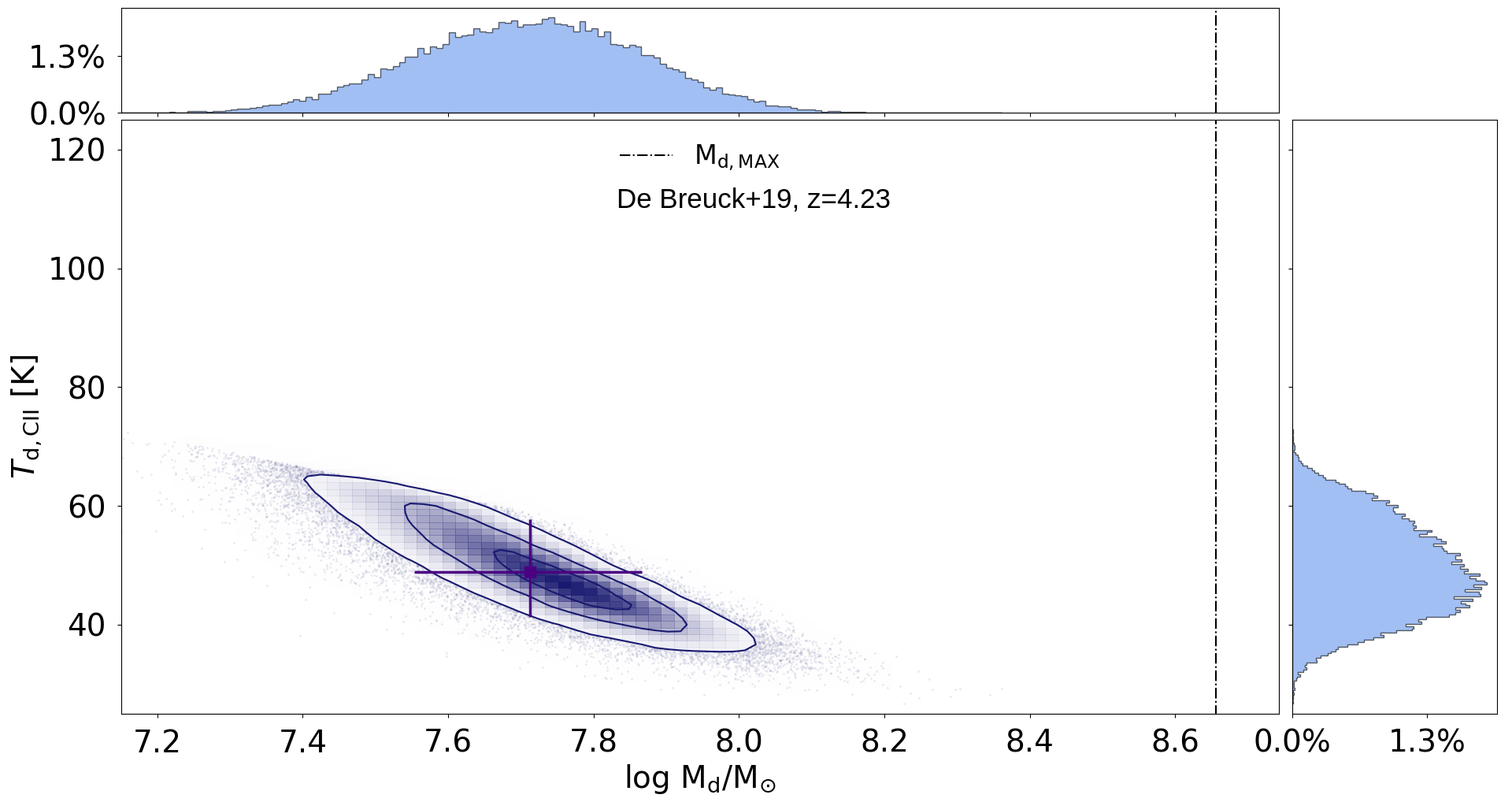}
    \includegraphics[width=0.8\linewidth]{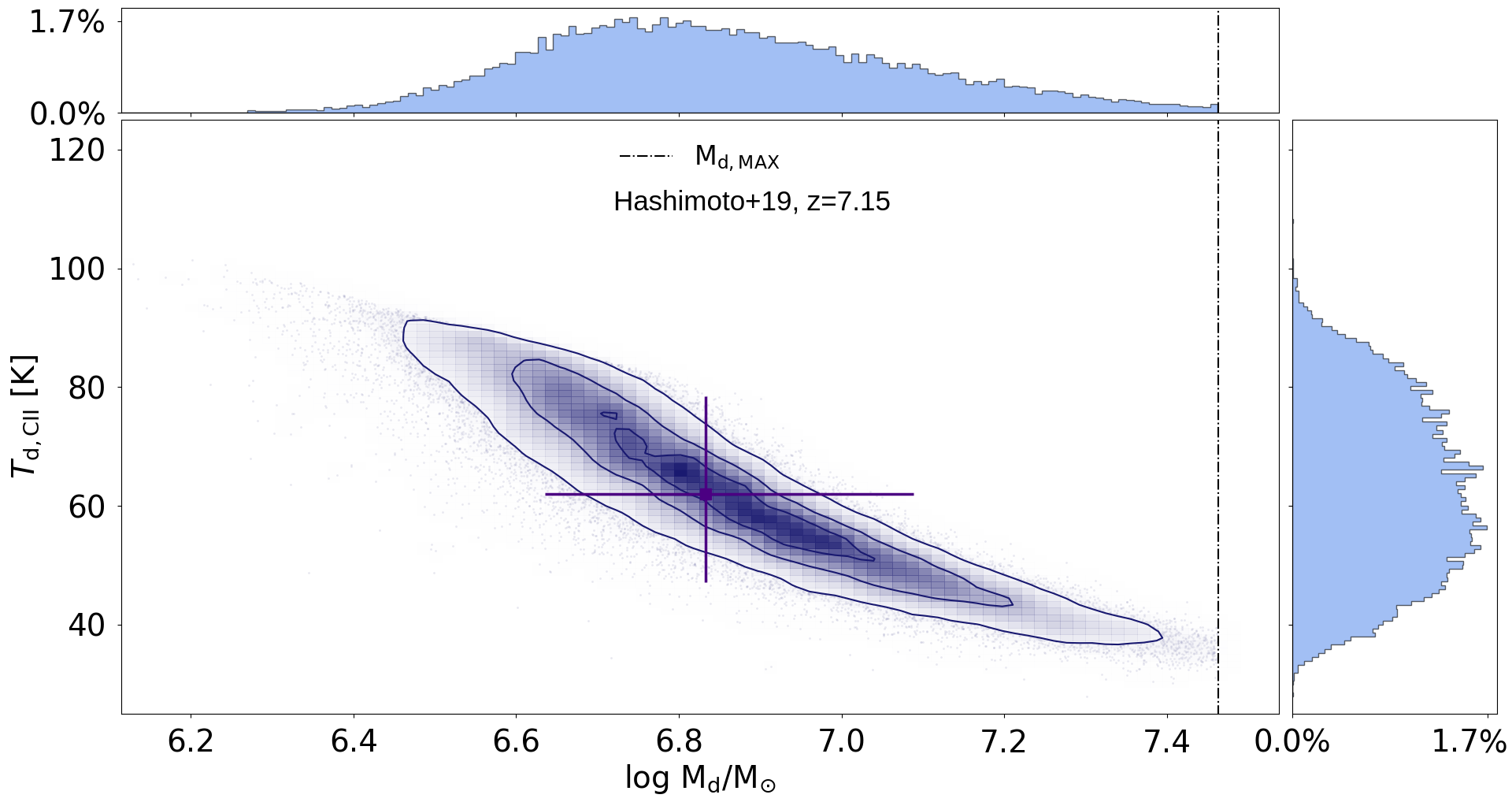}
    \includegraphics[width=0.8\linewidth]{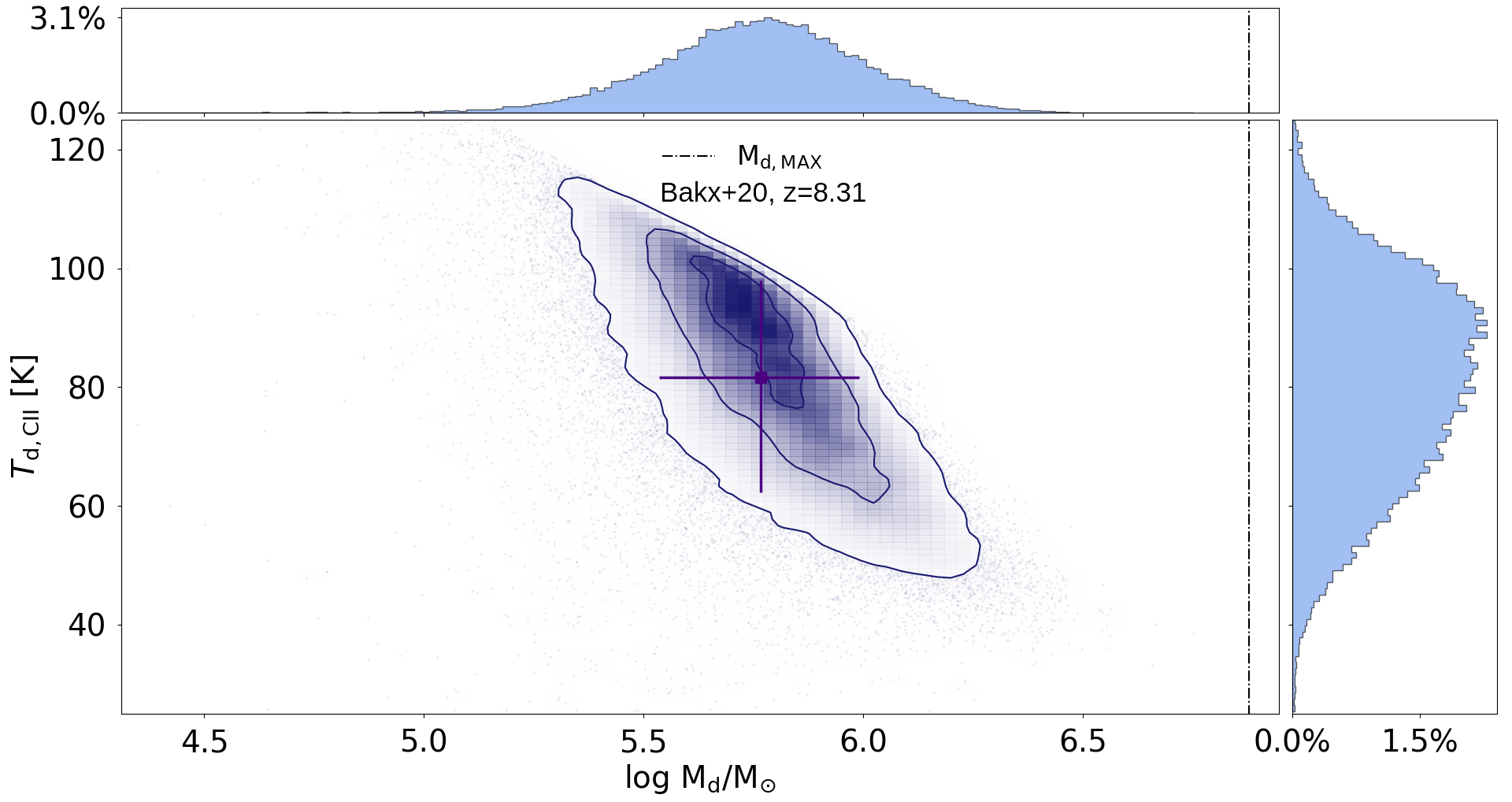}
    \caption{Recovered distribution of $T_{\rm d, CII}$ as a function of $M_{\rm d}$ for the galaxies in our high-$z$ template sample (same order as in Tab. \ref{tabhz}). The contours show the $(16,50,84)$ percentiles of the distribution. 
    The median value is represented by the purple square (alongside its 16 and 84 percentiles marked by the error bars). We also show the temperature (right) and mass (top) PDFs. The upper limit on the dust mass, computed through eq. \ref{mdmax}, is shown by the vertical black dot-dashed line in each panel. }
    \label{bakx}
\end{figure*}

\begin{table}
  \begin{center}
    \begin{tabular}{l|c|c|c|c|r}
        \hline
        Galaxy  & $\alpha_{\rm CII,hz}$ & $\mathrm{T_{\rm d, SED}}$ & $T_{\rm d, CII}$ & $M_{\rm d}$ & $y_{\rm d}$ \\
        &    & $ [\mathrm{K}]$& $ [\mathrm{K}]$ & $[10^{6}\ \mathrm{M_{\odot}]}$ &  $\mathrm{[M_{\odot}/SN]}$\\
        \hline\hline
        SPT0418-47$^{a}$ & $7\pm1$ & $45-58$ & $-$ &$110^{+10}_{-20}$ & $-$ \\\\
        \textbf{This work} & $5^{+2}_{-1}$ & $-$ & $49^{+9}_{-8}$ & $52^{+22}_{-16}$ & $0.2 \pm 0.1$ \\\\
         \hline
        
        B1465666$^{b}$ & $-$ & $48-61$ & $-$ & $6.5-13.2$ & $-$ \\\\
        \textbf{This work} & $2^{+2}_{-1}$& $-$  & $61^{+16}_{-15}$ & $6.8^{+5.4}_{-2.4}$ & $0.5^{+0.3}_{-0.1}$ \\\\
         \hline
        
        MACS0416Y1$^{c}$ & $-$ & $\ge 80$& $-$  & $0.2-0.6$ & $-$ \\\\
        \textbf{This work} & $2^{+2}_{-1}$ & $-$& $\le 82^{+16}_{-19}$ & $0.6^{+0.4}_{-0.2}$ & $0.2 \pm 0.1$\\
         \hline
        
    \end{tabular}
    \caption{Comparison between predicted (\quotes{This work}) and published properties of galaxies in our high-$z$ sample. We underline that the dust temperature estimates taken from the literature are obtained through SED fitting, while our predictions correspond to $T_{\rm d, CII}$. \textbf{References}:$^{a}$\citet{strandet2016redshift,2017BothwellSPT,2019A&A...631A.167D,ReuterSPT,2020Rizzo}. Here we show the intrinsic values, which are obtained by dividing by the magnification factor of the source $\mu = 32.7$ \citep{2019A&A...631A.167D}; $^{b}$\citet{big3drag}, and $^{c}$\citet{Bakx20}.}
     \label{tabhz1} 
  \end{center}
\end{table}

\subsection{Dust mass}
Our method also provides reliable estimates of the dust masses of distant galaxies. Once $\alpha_{\rm CII}$ has been determined, it is straightforward to derive $M_{\rm d}$ using eq. \ref{zanella}. 
As already discussed, we discard the solutions for which $M_{\rm d}>M_{\rm d, max}$, which is computed in eq. \ref{mdmax}. 

For SPT0418-47 we find $M_{\rm d}=0.5^{+0.2}_{-0.1}\ \times 10^{8}\ \mathrm{M_{\odot}}$, a value pretty consistent with that obtained from SED fitting $M_{\rm d}=1.1^{+0.1}_{-0.2}\ \times 10^{8}\ \mathrm{M_{\odot}}$. 
For B14-65666 we find $M_{\rm d}=6.8^{+5.4}_{-2.4}\ \times\ 10^{6}\ \mathrm{M_{\odot}}$. This is also consistent with the result obtained from SED fitting by \cite{big3drag}. They find $6.5<M_{\rm d}/10^{6}\ \mathrm{M_{\odot}}<13.2$ with $1.5 \le\beta\le 2$. 
In the case of MACS0416-Y1 we find $M_{\rm d}=0.6_{-0.2}^{+0.4}\ \times\ 10^{6}\ \mathrm{M_{\odot}}$. This is consistent with the result obtained from SED fitting by \cite{Bakx20}. They find $M_{\rm d}=0.2-0.6\ \times\ 10^{6}\ \mathrm{M_{\odot}}$ for $70\ \mathrm{K}<T_{\rm d, SED}<130\ \mathrm{K}$, and $\beta=2$. 

We also compute the dust yield per SN, $y_d$, required to produce the above dust masses, which are consistent with previous estimates in the literature. We use the formula in \cite{nostro}:
$y_d = M_{\rm d}/M_{\star}\nu_{\rm SN}$, where $\nu_{\rm SN} = (53\ \mathrm{M_{\odot}})^{-1}$ is the number of SNe per solar mass of stars formed \citep{10.1046/j.1365-8711.2000.03209.x}.
For all the three sources it is $y_d \simlt 1\ \mathrm{M_\odot}$ (see Tab. \ref{tabhz1} for the value of $y_d$ in each galaxy), i.e. within the allowed range given in the latest SNe dust production studies by \cite{lesniewska2019dust}. They find that up to $y_d \le 1.1\ \mathrm{M_{\odot}}$ per SN can be produced, where the exact value depends on the amount of dust which is destroyed during the explosion ($1.1\ \mathrm{M_{\odot}}$ corresponds to the extreme case of no dust destruction). 

The presence of warmer dust ($T_{\rm d,SED} \simgt 50\ \mathrm{K}$) in these high-$z$ sources alleviates the large dust mass requirements set by the observed FIR luminosity. This is particularly relevant in the context of early galaxies. Allowing for lower dust masses prevents from invoking super efficient dust production by stellar sources, which is difficult to reconcile with both data and theoretical models \citep[for a detailed discussion see e.g.][]{nostro}. 

\section{Molecular gas content}\label{alfamolsec}
Besides providing a reliable determination of the dust temperature, our method offers a physical interpretation of the \citet{zanita19} relation. 
To show this, we parallel the analysis in Sec. \ref{alfaCIISFR}. Here we substitute the KS relation with the following expression linking the star formation and molecular gas surface density $\Sigma_{\rm H_{2}}$ \citep{Krum}\footnote{We adopt the standard units used for these quantities:  $\Sigma_{\rm SFR}\ [\mathrm{M_{\odot}\, kpc^{-2}\,yr^{-1}}]$, $t_{\rm depl}\ \mathrm{[Gyr]}$, and in eq. \ref{alfamol}, $\alpha_{\rm CII, mol}\ \mathrm{[M_{\odot}/L_{\odot}]}$.}:
\begin{equation}\label{tdepl}
    \Sigma_{\rm SFR}  = 10^{-9} \frac{\Sigma_{\rm H_{2}}}{t_{\rm depl}}.
\end{equation}
where $t_{\rm depl}$ is the depletion time. Combining eq. \ref{tdepl} with the DL relation and the definition of the molecular conversion factor, $\alpha_{\rm CII, mol} = \Sigma_{\rm H_{2}}/\Sigma_{\rm CII}$, we find \begin{equation}\label{alfamol}
    \alpha_{\rm CII, mol}= \frac{t_{\rm depl}}{3.3 \times 10^{-2}} \Sigma_{\rm SFR}^{-0.075}.
\end{equation}
The dependence of $\alpha_{\rm CII, mol}$ on $\Sigma_{\rm SFR}$ is extremely weak, in contrast with $\alpha_{\rm CII} \propto \Sigma_{SFR}^{-0.3}$ (total gas conversion coefficient\footnote{The exponent $-0.3$ is an average value between the $-0.36$ and $-0.29$ found in eq. \ref{alfaKS}, \ref{alfahz}.}, see Sec. \ref{alfaCIISFR}). We can understand this result in physical terms as both $\mathrm{H_{\rm 2}}$ and \CII\ emission trace closely ongoing star formation. Since both $\Sigma_{\rm H_{2}}$ and $\Sigma_{\rm CII}$ scale almost linearly with $\Sigma_{\rm SFR}$, their ratio is virtually independent of this quantity. Instead, the total gas reservoir is less sensitive to star formation (see eq. \ref{KSR}). Therefore in the ratio $\Sigma_{\rm gas}/\Sigma_{\rm CII}$ the dependence on $\Sigma_{\rm SFR}$ does not cancel out.

Most recent results by \cite{walter2020evolution} suggest that $t_{\rm depl}$ is nearly constant above redshift $z>2$, and then increases slightly from $t_{\rm depl}\sim 0.4\ \mathrm{Gyr}$ at $z\sim 2$, to $t_{\rm depl}\sim 0.7\ \mathrm{Gyr}$ at $z=0$. Substituting these values in eq. \ref{alfamol}, we find $\alpha_{\rm CII, mol}= (12-21)\, \Sigma_{\rm SFR}^{-0.075}$. This result is compatible with the measurement of $M_{\rm H_{2}}/L_{\rm CII} = 31^{+31}_{-16}\ \mathrm{M_{ \odot}/L_{\odot}}$ derived by \cite{zanita19} in a sample of galaxies at $z \sim 0-6$. Recently, \cite{dessaugeszavadsky2020alpinealma} found this $M_{\rm H_{2}}/L_{\rm CII}$ ratio to hold also in the \CII-detected galaxies at $z \sim 4-6$ targeted by the ALPINE survey, albeit with some uncertainties. \footnote{More precisely, \cite{dessaugeszavadsky2020alpinealma} find a good agreement between molecular gas masses derived from \CII\ luminosities (using the relation by \citealt{zanita19}), dynamical masses, and rest-frame $850\ \mathrm{\mu m}$ luminosities (extrapolated from the rest-frame $158\ \mathrm{\mu m}$ continuum).}.

On average, previous works indicated longer depletion times $t_{\rm depl} \sim [0.5,2]\ \mathrm{Gyr}$ both in local and high-$z$ galaxies \citep[$z \sim 6-0$, see e.g.][]{2008AJ....136.2846B,genzel2010,Daddi_2010,2013AJ....146...19L,2014ApJ...793...19S,2015ApJ...800...20G,2015A&A...573A.113B,2015A&A...577A..50D,Schinnerer_2016,Scoville_2017,Saintonge_2017,dessaugeszavadsky2020alpinealma}. Nevertheless, the observed scatter in $t_{\rm depl}$ is within measurement errors by \citealt{zanita19}. The variation of $\alpha_{\rm CII, mol}$ is significantly smaller than that of $\alpha_{\rm CII}$. Already within our limited sample of $23$ galaxies, $\alpha_{\rm CII}$ varies by nearly two orders of magnitude due to its strong dependence on $\Sigma_{\rm SFR}$ and $\kappa_s$ (see Fig. \ref{KS}).

\subsection{Molecular gas fraction}

Armed with the expressions for $\alpha_{\rm CII}$ (eq. \ref{alfaKS}) and $\alpha_{\rm CII, mol}$ (eq. \ref{alfamol}) we intend to study the redshift evolution of the ratio $\alpha_{\rm CII, mol}/\alpha_{\rm CII}= \Sigma_{\rm H_2} / \Sigma_{\rm gas}$. To this aim, since $\alpha_{\rm CII}\propto \Sigma_{\rm SFR}^{-0.3}$, we need to provide a qualitative prescription for the redshift evolution of the average $\bar{\Sigma}_{\rm SFR}(z) $ in normal galaxies. 

\begin{figure*}
    \centering
    \includegraphics[width=1.0\linewidth]{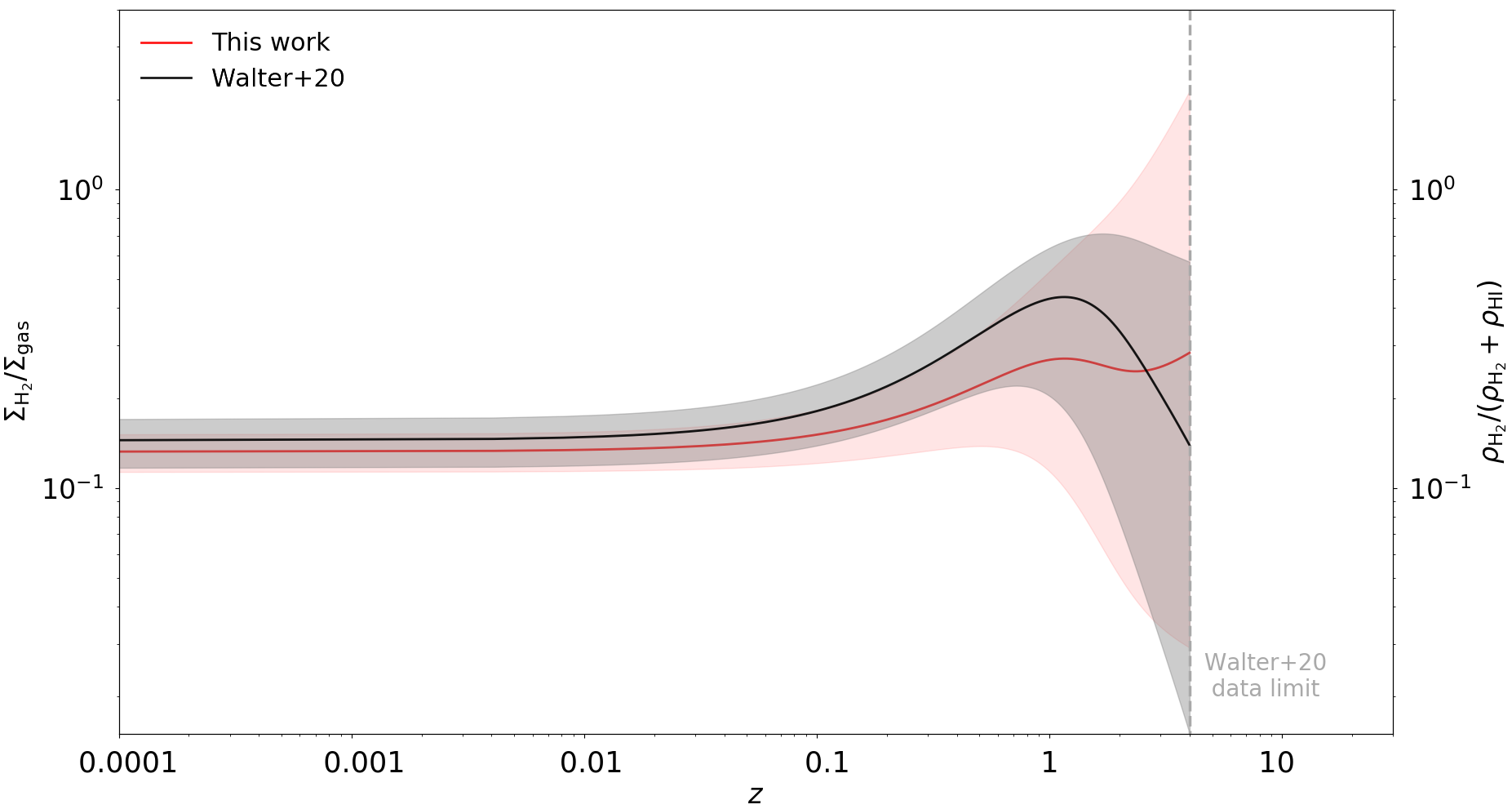}
    \caption{Redshift evolution of the molecular gas fraction $f_{\rm H_2}(z)=\Sigma_{\rm H_2}/\Sigma_{\rm gas}$. The red line represents our fiducial estimate. The black line shows the $\rho_{\rm H_2}/(\rho_{\rm H_2}+\rho_{\rm HI})$ trend observed by  \citealt{walter2020evolution}; the vertical grey dashed line refers to the highest redshift considered in their analysis. For sake of the comparison here we consider the same $t_{\rm  depl}(z)$ as in \citealt{walter2020evolution}.}
    \label{fmol}
\end{figure*} 

We consider the cosmic SFR comoving density, $\psi$, derived by \cite{Madau_2014} in the range $z=0.01-8$. We combine $\psi$ with the evolution of the effective radius, $r_\star \approx r_{\rm e}=6.9\times (1 + z)^{1.2}\ \mathrm{kpc}$, derived by \cite{Shibuya_2015} for a HST sample of $\sim 19,000$ galaxies at $z=0-10$ to obtain $\bar{\Sigma}_{\rm SFR}(z) = {\psi}/{\pi r_{\rm e}^2}$, and the corresponding expression for $\alpha_{\rm CII}(z)$ from eq. \ref{alfaKS}. 

For simplicity, in computing $\alpha_{\rm CII}(z)$ we consider $\kappa_s=1$ as on average we expect most local and low-$z$ galaxies to lie on the KS-relation\footnote{If $\kappa_s>1$ the $\Sigma_{\rm H_2} / \Sigma_{\rm gas}$ curve is shifted upwards, as we are reducing $ \Sigma_{\rm gas}\propto \alpha_{\rm CII} \propto \kappa_s^{-5/7}$, without affecting $\Sigma_{\rm H_2}$. Hence, at higher redshift deviations are expected to occur due to the burstiness of galaxies.}.  
In parallel to the result shown in Sec. \ref{alfamolsec}, we write $\alpha_{\rm CII, mol}(z) = (12-21)\, \bar{\Sigma}_{\rm SFR}^{-0.075}$. We can then compute the ratio: 
\begin{equation}
f_{\rm H_2}(z) \equiv \Sigma_{\rm H_2} / \Sigma_{\rm gas} =\alpha_{\rm CII, mol}/\alpha_{\rm CII} \sim \left(\frac{\bar{\Sigma}_{\rm SFR}}{\mathrm{M_{\odot}kpc^{-2}yr^{-1}}}\right)^{0.225}   
\end{equation}
The redshift evolution of the molecular fraction in galaxies has been experimentally determined by \cite{walter2020evolution} from observations\footnote{The \HI density is obtained by combining measurements of \HI emission in the local universe \citep[see e.g.][]{zwaan05} with quasar absorption lines at higher $z$ \citep[see e.g.][]{2009ApJ...696.1543P}; $\rho_{\rm H_2}(z)$ determination is based on CO and FIR dust continuum data \citep[e.g. reviews by][]{2013carilli,Tacconi_20_rev,2020ARA&A..5821820P,hodge2020highredshift}.} of molecular, $\rho_{\rm H_2}$, and atomic, $\rho_{\rm HI}$, gas densities at $z\simlt 4$.  In Fig. \ref{fmol} we show the comparison of $f_{\rm H_2}(z)$ with the observed redshift evolution of of the empirical $\rho_{\rm H_2}/(\rho_{\rm HI}+\rho_{\rm H_2})$ ratio. 

The two approaches yield a pretty consistent evolution trend, albeit they are both affected by large uncertainties.  We find that on average  $f_{\rm H_2}(z)$ increases by a factor of $\sim 2$ from $z=0.01$ to $z=1$, in agreement with the trend found by \cite{walter2020evolution} ($\sim 1.7-3.6$). However, at $z>1$ the two trends might be different, as we predict a possible further increase in $f_{\rm H_2}(z)$. This can be due to (a) a further increase in the $M_{\rm H_2}/M_{\rm gas}$, and/or (b) an increase in the ratio $r_{\rm gas}/r_{\rm H_2}$ ratio. The first case seems to be disfavoured by theoretical studies (see e.g. \citealt{2017MNRAS.467..115D}), as both the H$_2$ and \HI evolution become steeper with redshift at fixed stellar mass. The second possibility is instead suggested by recent works showing the presence of \CII\ emission at high-$z$ around $\times 1.5-3$ times more extended than the stellar (and possibly molecular) mass \citep[see e.g.][]{ carniani:2017oiii,carniani2018clumps,2019ApJ...887..107F,2020arXiv200300013F,2020A&A...633A..90G,Carniani20}. Clarifying this uncertainty is crucial as the assumption that $f_{\rm H_2}\approx 1$ at high-$z$ is widely used to derive molecular gas masses from dynamical \citep{Daddi_2010,genzel2010,dessaugeszavadsky2020alpinealma} and dust  \citep{scoville2016ism,dessaugeszavadsky2020alpinealma} masses.

\section{Summary and Conclusions}\label{summary}

We have proposed a novel method to derive the dust temperature in galaxies, based on the combination of continuum and \CII\ line emission measurements, which breaks the SED fitting degeneracy between dust mass and temperature. The method allows constraining $T_d$ from a single band observation at $1900\ \mathrm{GHz}$ (rest-frame). We conveniently provide analytic expressions in eq. \ref{td1pto} for a direct application. 

Besides, the same method offers a physical explanation for the empirical relation found by  \citet{zanita19} between \CII\ luminosity and molecular gas. We also derive the relation between total gas surface density and \CII\ surface brightness, $\Sigma_{\rm gas}=\alpha_{\rm CII}\Sigma_{\rm CII}$. 
By combining such relations we predict the redshift evolution of the molecular gas fraction defined here as $\Sigma_{\rm H_2} / \Sigma_{\rm gas}$. 

We summarise our main findings below:
\begin{itemize}
    \item \textbf{Dust temperature from \CII\ data at high-$z$}: using a single band observation, with our method, we can constrain the dust temperature as well as with the commonly used SED fitting in multiple bands. We recover dust temperatures consistent with literature data (within $1\sigma$) out to redshift $z=8.31$;  
    
    \item \textbf{Gas-to-\CII\ luminosity relation}: the total gas conversion coefficient $\alpha_{\rm CII}$ strongly depends on the SFR surface density ($\sim \Sigma_{\rm SFR}^{1/3}$) and the burstiness of galaxies (see eq. \ref{alfaKS}). When computing the analogous conversion factor for the molecular gas $\alpha_{\rm CII, mol}$, we find that the dependence on $\Sigma_{\rm SFR}$ nearly cancels out, hence $\alpha_{\rm CII,mol}\approx $const. (see eq. \ref{alfamol}); 
    
    \item \textbf{Molecular gas fraction}: we find that $f_{\rm H_2}(z)$ on average increases with $z$ by a factor $\approx 2$ from $z=0.01$ to $z=1$. This is consistent with the trend observed by \cite{walter2020evolution}. We predict a possible further increase at $z>1$. This could be caused by a rise of the $\mathrm{H_2}$ content, and/or a change in the relative extension of $\mathrm{H_2}$ and \HI gas.
\end{itemize}

Assuming a dust temperature, as usually done in high-$z$ galaxy observations analysis, introduces large uncertainties on the derived dust masses, infrared luminosities, and star formation rates (see also e.g. \cite{nostro} for a detailed discussion). Our method can improve the reliability of the interpretation of \CII\ and continuum observations from ALMA and NOEMA. This is particularly relevant in the context of recent ALMA large programs targeting \CII\ emitters at high-$z$, such as ALPINE \citep{lafev_alpine,alpineCII_SFR,bethermin_alpine,Schaerer_alpine}, REBELS (PI: Bouwens), and others. With future instruments such as JWST, providing more accurate metallicity measurements, it will be possible to improve current estimates of the dust-to-gas ratios at high-$z$. This will further enhance the precision of our dust temperature determinations.

\section*{Acknowledgements}
LS, AF, SC, AP, LV acknowledge support from the ERC Advanced Grant INTERSTELLAR H2020/740120 (PI: Ferrara). Any dissemination of results must indicate that it reflects only the author’s view and that the Commission is not responsible for any use that may be made of the information it contains. Partial support from the Carl Friedrich von Siemens-Forschungspreis der Alexander von Humboldt-Stiftung Research Award is kindly acknowledged (AF). We thank R. Kennicutt, T. Díaz-Santos, and C. De Breuck for the useful discussions, and for the help in retrieving the data from their surveys, respectively KINGFISH, GOALS, and SPT. 

\section{DATA AVAILABILITY}

Part of the data underlying this article were accessed from the computational resources available to the Cosmology Group at Scuola Nor-
male Superiore, Pisa (IT). The derived data generated in this research
will be shared on reasonable request to the corresponding author.

\appendix \label{appdL}

\section{Hints from simulations}\label{appendix}
\begin{figure*}
    \centering
    \includegraphics[width=1.0\linewidth]{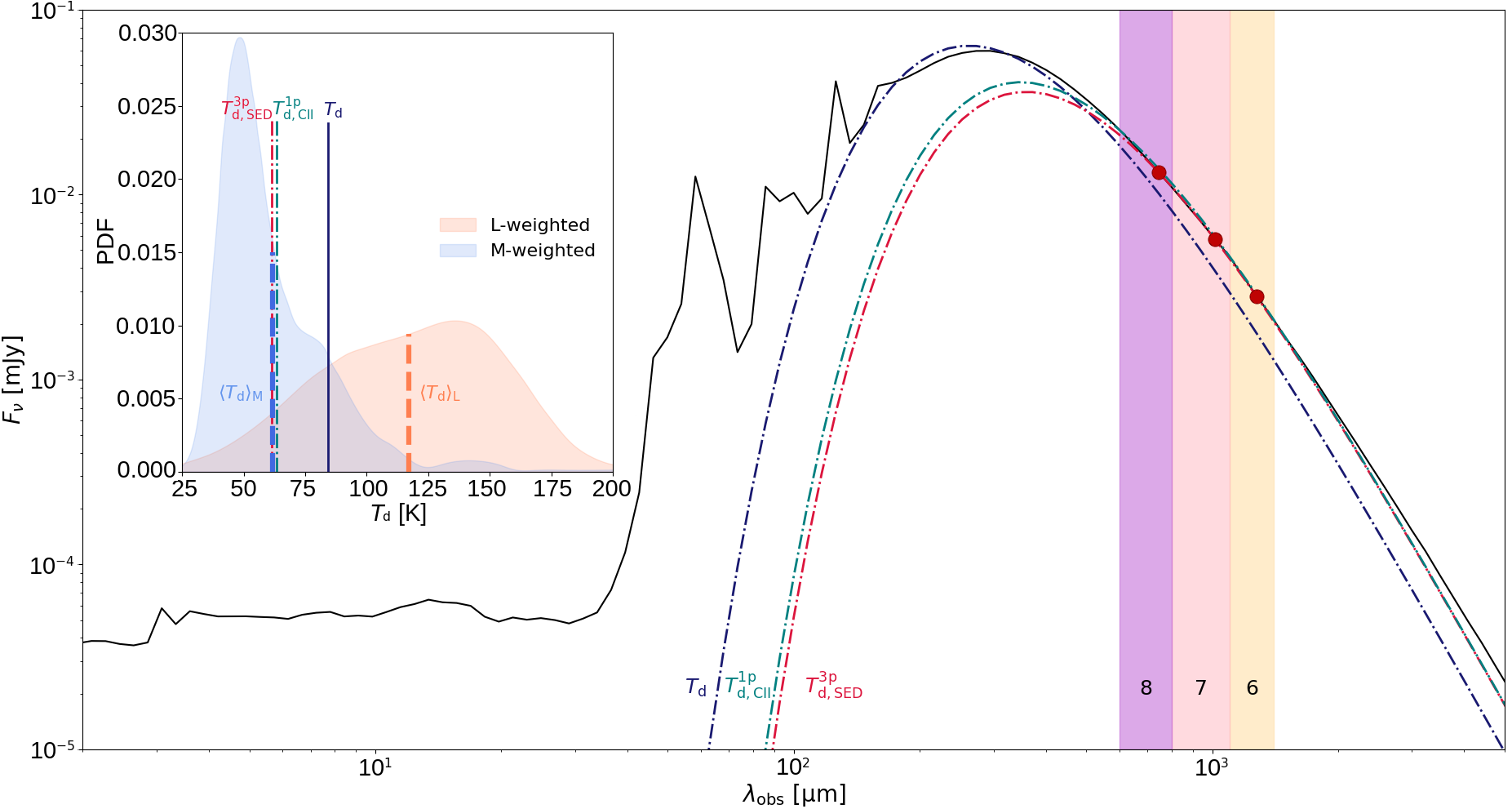}
    \caption{SED for the simulated galaxy Zinnia (serra05:s46:h0643, black solid line) extracted from the SERRA simulation suite. The dotted-dashed lines show the curves obtained through a single temperature grey body fitting of the SED, with the two following methods: (a) a canonical SED fitting performed considering the three red \quotes{data points} in ALMA bands $6,7,8$ (red line, $T^{\rm 3p}_{\rm d, SED}$); (b) same as (a) but considering the full (i.e. MIR and FIR) galaxy SED (blue line, $T_{\rm d}$); (c) the method presented in this work; it uses a single continuum observation and the \CII\ emission (green line, $T^{1p}_{\rm d, CII}$). The shaded regions mark the ALMA bands $6$ to $8$. \textit{Subplot}: comparison among the above dust temperatures values, the luminosity- (orange), and mass-weighted (blue) dust temperature PDFs derived for Zinnia. The PDFs mean values  ($\left< T_{\rm d} \right>_{\rm L}$ and $\left< T_{\rm d} \right>_{\rm M}$) are indicated by dashed lines. See text for a detailed discussion. }
    \label{simul}
\end{figure*}
The method developed in this paper can reliably determine the dust temperature in a galaxy for which only a \textit{single} simultaneous observation for the \CII\, line and underlying continuum is available. This value corresponds to the canonical temperature, $T_{\rm d, SED}$, that one would normally define from fitting the SED with a single temperature grey body formula.  As already mentioned, $T_{\rm d, SED}$ does not necessarily correspond to the physical dust temperature which instead is distributed according to a given Probability Distribution Function \citep[PDF, see e.g.][Di Mascia et al. in prep.]{behrens2018dusty,nostro}. Hence, it is instructive to understand the relation among  $T_{\rm d, SED}$ and the PDF properties.

In the context of theoretical studies, i.e. both analytical models and simulations, the dust temperature PDF is actually available. Various weighting procedures can be applied to this PDF, and then the average can be compared to the observational results. 
In particular, the most commonly adopted are the mass- (M-weighted) and luminosity-weighted (L-weighted; $L \propto M_{\rm d} T_{\rm d}^{4+\beta}$). The M-weighted temperature traces the most abundant cold temperature component; instead, the L-weighted is biased towards hotter but less massive dust component present in star forming regions where it is efficiently heated by the UV emission from newborn stars, see e.g. \citealt[][Di Mascia et al. in prep.]{behrens2018dusty,nostro}). Neither of them is traced by $T_{\rm d, SED}$. Indeed cold dust nearly in equilibrium with the CMB is not observable in emission; hot dust (if present) emits mainly in the MIR, where it is largely responsible for distortions of the single temperature grey body (see e.g. \citealt{2012casey,casey2018}). 

At high-$z$ such distortion is not observable, as only the long-wavelength part of the SED spectra is currently accessible with ALMA (bands 6,7, and 8). However, locally, where the whole SED is well sampled, it has been observed and studied by e.g. \cite{2012casey} within $z\sim 0$ sub-millimetre galaxies. In light of these considerations, the most appropriate and clean choice when comparing theoretical results with observations is to perform a single temperature grey-body fit to the simulated SEDs in order to consistently obtain $T_{\rm d, SED}$. In Fig. \ref{simul} we show the result of applying this procedure to the SED of the simulated $z\sim 6.7$ galaxy Zinnia (a.k.a. serra05:s46:h0643) from the SERRA simulation suite.

Full details on SERRA simulations are given in Pallottini in prep. and can be summarized as follows. Simulations zoom in on the evolution of $M_{\star} \sim 10^{10}\ \mathrm{M_{\odot}}$ galaxies from $z=100$ to $z=6$ with a mass (spatial) resolution of the order of $10^4\ \mathrm{M_{\odot}}$ ($30$ pc at $z=6$)\footnote{The simulation adopts a multi-group radiative transfer version of the hydrodynamical code RAMSES \citep{2013MNRAS.429.3068T,2013MNRAS.436.2188R} that includes thermochemical evolution via KROME (\citealt{2014MNRAS.439.2386G}, \citealt[see][for the network and included processes]{2014MNRAS.441.2181B,pallottini:2017}), which is coupled to the evolution of radiation \citep{pallottini:2019,2020MNRAS.497.4718D}. Stellar feedback includes SN explosions, OB/AGB winds, and both in the thermal and turbulent form \citep[see][for details]{pallottini:17a}.}.
\CII\ emission is obtained by post-processing using grids of CLOUDY \citep{2017RMxAA..53..385F} models accounting for the internal structure of molecular clouds \citep{vallini2017molecular,pallottini:2019}. Additionally, SKIRT \citep{2015A&C....12...33B,2015A&C.....9...20C} is used to obtain UV and dust continuum emission, with a setup similar to \cite{behrens2018dusty}.

\begin{table*}
  \begin{center}
    \begin{tabular}{l|c|c|c|c|c|c|c|r}
        \hline
        Galaxy & $z$ & $F_{\nu_0}$ & $Z$                   & $ \log \Sigma_{\rm SFR}$                   & $ L_{\rm CII}$                 & $\kappa_s$  & $y=r_{\rm CII}/r_{\star}$  & $M_{\star}$\\
               &     & $[\mathrm{\mu Jy}]$ & $[Z_{\odot}]$ & $\mathrm{[M_{\odot}\ yr^{-1}\ kpc^{-2}]}$  & $[10^{8}\ \mathrm{L_{\odot}]}$ &              &                       &         $[10^{9}\ \mathrm{M_{\odot}}]$  \\
        \hline \hline
     Zinnia &  $6.6847$ & $2.81$           & $0.07$        &  $2.56$                                   & $2.05$                         & $4.29$           & $1.00^{\star}$                 & $2.19$\\
        \hline

    \end{tabular}
    \caption{Properties of our high-$z$ simulated galaxy Zinnia (a.k.a. serra05:s46:h0643). We note that the parameter $y=r_{\rm CII}/r_{\rm \star}=1.0$ is selected by definition, i.e. we only consider the emission coming from the central $\sim 1.5\ \mathrm{kpc}$ region. }
     \label{serratab} 
  \end{center}
\end{table*}

\begin{table}
  \begin{center}
    \begin{tabular}{l|c|c|c|r}
        \hline
        Galaxy  & $\alpha_{\rm CII,hz}$ & $\mathrm{T_{\rm d, SED}}$ & $M_{\rm d}$              & $y_{\rm d}$ \\
        &    & $ [\mathrm{K}]$ & $[10^{5}\ \mathrm{M_{\odot}]}$                                &  $\mathrm{[10^{-3}\ M_{\odot}/SN]}$\\
        \hline\hline
        
         serra0643 & $8.9$                 & $62 \pm 2$            & $2.0\pm 0.8$              & - \\\\
        \textbf{This work} & $8.76\pm0.07$ & $63.4\pm 0.5$       & $1.94\pm 0.03$       & $4.67 \pm 0.06$ \\\\
         \hline
        
    \end{tabular}
    \caption{Comparison between the properties predicted with our method (\quotes{This work}), and derived through a single temperature grey body fitting of the simulated flux in ALMA band 6,7,8 of galaxy serra05:s46:h0643. In the SED fitting procedure, we keep the dust emissivity index fixed at $\beta=2.0$, as in our analytical method, and consider a $1\%$ uncertainty on all the galaxy properties derived from the simulation and listed in Tab. \ref{serratab}. We underline that our predictions correspond to $T_{\rm d, CII}$.}
     \label{resultsimul} 
  \end{center}
\end{table}

The main properties of Zinnia are summarised in Tab. \ref{serratab}. We proceed to compute and compare the following temperatures:
\begin{itemize}
    \item $T_{\rm d}$: dust temperature obtained from fitting the full (i.e. MIR and FIR) galaxy SED with a single-T grey-body;
    \item $T^{\rm 3p}_{\rm d, SED}$: dust temperature obtained from fitting the galaxy SED at the frequencies corresponding to ALMA band $6,7,8$ with a single-T grey-body;
    \item $T^{1p}_{\rm d, CII}$: dust temperature obtained with our method combining a single continuum data at $1900\ \mathrm{GHz}$ (rest-frame) with the \CII\, line emission data, as described in Sec. \ref{hztemplate};
    \item $\left< T_{\rm d} \right>_{\rm M}$: M-weighted dust temperature;
    \item $\left<  T_{\rm d} \right>_{\rm L}$: L-weighted dust temperature.
\end{itemize}
We underline that in all these computations, as in the rest of the paper, we keep the dust emissivity index fixed to $\beta_{\rm d}=2.0$. Such an assumption is reasonable as this is close to the emissivity index retrieved from the simulation ($\beta_{\rm d}=1.7-2$, see \cite{behrens2018dusty} for the Radiative Transfer details). Hence, the free parameters in the fitting procedure are the dust temperature and dust mass ($40\ \mathrm{K} \le T_{\rm d} \le 200\ \mathrm{K}$, and $ 10^3\ \mathrm{M_{\odot}} \le M_{\rm d} \le 10^7\ \mathrm{M_{\odot}}$). 

All these temperatures are compared in the subplot in the upper left corner of Fig. \ref{simul}. Our method gives a dust temperature value $T^{1p}_{\rm d, CII} = 63.4\pm 0.5\ \mathrm{K}$, consistent with the result that one obtains with the usual SED-fitting technique using three points corresponding to the available ALMA bands at this redshift, $T^{\rm 3p}_{\rm d, SED}=62\pm 2\ \mathrm{K}$. For this galaxy, the value of $T^{\rm 3p}_{\rm d, SED} \sim T^{1p}_{\rm d, CII}$ is also consistent with the M-weighted temperature, $\left< T_{\rm d} \right>_{\rm M} = 61\ \mathrm{K}$. Instead both $\left<  T_{\rm d} \right>_{\rm L}=117$ K and $T_{\rm d}=84$ K are larger than the previous values as they are more sensitive to the small amount of dust with physical temperatures up to $150\ \mathrm{K}$ (see the L-weighted PDF in the subplot of Fig. \ref{simul}). 

This comparison shows that the single-T approximation often used might lead to a misinterpretation of the physical properties of the galaxy depending directly on $T_{\rm d}$.  
Moreover, whenever theoretical studies and observations are compared, it is necessary to pay particular attention to the definition of the dust temperature used and to the fitting procedure.  \textit{We suggest that a uniform, meaningful comparison is best performed using either $T^{\rm 3p}_{\rm d, SED}$ or, as we propose here, $T^{1p}_{\rm d, CII}$, when only a single measurement is available.} It is very reassuring that the two procedures yield essentially the same result. These quantities can be also easily derived from the simulated spectrum, and readily compared with data.

\bibliographystyle{mnras}
\bibliography{latest}

\bsp	

\label{lastpage}
\end{document}